\documentclass[11pt]{article}
\usepackage{natbib}

\usepackage{siunitx}

\usepackage{lineno,hyperref,amsmath,graphics,dcolumn}
\usepackage{tabularx}
\usepackage{array}
\usepackage{adjustbox}
\usepackage{multirow} 
\usepackage{subfig}
\usepackage{caption}
\usepackage{graphicx}
\usepackage{xargs}                      
\usepackage{arydshln}
\usepackage[colorinlistoftodos,prependcaption,textsize=tiny]{todonotes}
\newcommandx{\sara}[2][1=]{\todo[linecolor=red,backgroundcolor=red!25,bordercolor=red,#1]{#2}}
\newcommandx{\giovanna}[2][1=]{\todo[linecolor=blue,backgroundcolor=blue!25,bordercolor=blue,#1]{#2}}
\newcommandx{\daniela}[2][1=]{\todo[linecolor=green,backgroundcolor=green!25,bordercolor=green,#1]{#2}}
\newcommandx{\stefano}[2][1=]{\todo[linecolor=Plum,backgroundcolor=Plum!25,bordercolor=Plum,#1]{#2}}
\newcommandx{\alessandro}[2][1=]{\todo[disable,#1]{#2}}
\makeatletter
\let\@fnsymbol\@arabic
\makeatother

\title{
Integration of presence-only data from several sources. A case study on dolphins' spatial distribution
}
\author{Sara Martino\thanks{Department of Mathematical Sciences, NTNU, Norway}, Daniela Silvia Pace$^{*}$\thanks{Department of environmental biology, Sapienza University of Rome, Italy, * corresponding author danielasilvia.pace@uniroma1.it}, Stefano Moro$^2$, Edoardo Casoli$^2$,\\ Daniele Ventura$^2$, Alessandro Frachea$^2$, Margherita Silvestri$^2$, \\ Antonella Arcangeli\thanks{ISPRA, Italian Institute for Environmental Protection and Research, Rome, Italy}, Giancarlo Giacomini$^2$,  Giandomenico Ardizzone$^2$, \\ Giovanna Jona Lasinio\thanks{Department of Statistical Sciences, Sapienza University of Rome, Italy}\\
}

\begin{document}

\maketitle

\begin{abstract}
\begin{itemize}
 \item Presence-only data are a typical occurrence in species distribution modelling. They include the presence locations and no information on the absence. Their modelling usually does not account for detection biases.
    \item In this work, we aim to merge three different sources of information to model the presence of marine mammals. The approach is fully general and it is applied to two species of dolphins in the Central Tyrrhenian Sea (Italy) as a case study. 
    \item Data come from the Italian Environmental Protection Agency (ISPRA) and Sapienza University of Rome research campaigns, and from a careful selection of social media (SM) images and videos.
    \item We build a Log Gaussian Cox process where different detection functions describe each data source. For the SM data, we analyse several choices that allow accounting for detection biases.
    \item Our findings allow for a correct understanding of \emph{Stenella coeruleoalba} and\emph{Tursiops truncatus} distribution in the study area. The results prove that the proposed approach is broadly applicable, it can be widely used, and it is easily implemented in the \texttt{R software} using INLA and inlabru. We provide examples' code with simulated data in the supplementary materials.
\end{itemize}
\end{abstract}


\section{Introduction}

Progress of ecological science is more and more reliant on combining data from diverse sources \citep{Fletcher2019}. This approach can increase the comprehension of ecological processes for both research and conservation purpose \citep{Paceetal2019}. Data availability to model species distribution, for example, is rapidly expanding thanks to the fast development of new technologies \citep{soranno2014macrosystems}, the growth of citizen science initiatives \citep{Florence2020.06.02.129536, Sicachaparada2020} and the opportunity of exploiting huge information harvested from the social media platforms \citep{mikula2016internet, Paceetal2019}. The latter data types can be intrinsically challenging to merge in with existing, valued and validated data collected via standard research protocols, yet if that can be achieved, they can offer enrichment of existing data to generate powerful insights and even reducing the costs of collecting data conventionally \citep{Bryman2018}. 
Nevertheless, heterogeneous data are complex to manage as they are 'polymorphic' in nature and affected by numerous forms of bias and limitations \citep{Isaac2015}. Information on species occurrence collected at sea by sea-users, for example, is characterised by a different spatiotemporal distribution of effort, which can be biased toward easily accessible habitat and times with better weather, or known areas of use \citep{Corkeron2011, Sicachaparada2020}. Hence, a simple data pooling \citep{Fletcher2019} with data gathered under conventional research methodologies is not enough to reliably model the presence of a species considering different explanatory variables – both environmental and anthropogenic – and to define its distribution over multiple spatial and temporal scales.

Integrated distribution modelling (IDM), i.e. the practice of fitting species distribution models with more than one observation practice \citep{Isaac2020}, is a new approach to combine different datasets, preserving the strengths of each and adjusting, at least to some extent, their limitations. IDM sets a spatial – or spatio-temporal – latent state, here statistically defined as a Point Process, of the sites where the animals were sighted, described by a series of covariates shared by different datasets. Multiple observation sub-models can be estimated from them, each describing a part of the latent state. Here we use the IDM approach on sighting data derived from different data sources (research, monitoring and social media) to predict the distribution of two dolphin species in the central Mediterranean Sea. The study of spatial distribution patterns of dolphin species is incredibly puzzling, as they spend much time under the water surface \citep{Redfern2006}, and establish whether they are present in a specific habitat involves a lot of visual/acoustic effort for scientists facing the constant background bias in the acquired data \citep{Redfern2017, breen2017}. 

Coping with several challenges, we propose a novel path to combine these different sources to provide cohesive summaries of the species' potential and realised distribution \citep{Isaac2020}.  First, as the available information is \textit{presence-only data}, we opt for Point Process as the most natural solution \citep[see][and references therein]{Miller2019}. Second, as several sources of bias are potentially present in the datasets, we propose models based on a location-dependent thinning of a Poisson process to reduce these biases \citep[see][and references therein]{Dorazio2014}; however, the parameters of these models are not fully identifiable unless the covariates of abundance are distinct and linearly independent of the covariates of detectability \citep{Dorazio2014}. In \citet{yuan2017} a flexible stochastic partial differential equation (SPDE) model describes the spatial structure  that is not accounted for by explanatory variables is proposed, and estimation is carried on using integrated nested Laplace approximation (INLA) in a Bayesian inference framework. The latter allows simultaneous fitting of detection and density models and permits prediction at an arbitrarily fine scale. Very recently \citet{Sicachaparada2020} adopt a similar approach using citizen science data on moose (\textit{Alces alces}) occurrence in Norway, accounting for the geographical bias (oversampling of "accessible" locations). For marine observations, the boat's size, the distance from the coast, policy regulations and weather conditions are just some of the factor that can affect the accessibility of an area. 
We aim to propose a new protocol for presence-only data fusion, where information sources include social media. We investigate several possible solutions and compare different types of detection function and accessibility explanations. We show how variation in the detection function affects ecological findings on two dolphin species with different spatial distribution. The approach is entirely broad and the selected species are representatives for different habitats. Hence they constitute a good benchmark for the entire proposal.  We provide R functions and example code to replicate our work in the online Supplementary material (\url{https://github.com/smar-git/SM-data-merging}).

\section{Materials and Methods}\label{sec:matmet}

\subsection{Study species}

Two dolphin species were selected for this study, the bottlenose (\emph{Tursiops truncatus}) and the striped dolphins (\emph{Stenella coeruleoalba}), both widely distributed throughout the Mediterranean Sea. The bottlenose dolphin is reported predominantly “coastal” or “inshore” \citep{Bearzi2012}, but its habitat changes depending on the region: it can inhabit shallow waters (less than 50 meters) close to the coast and at the mouths of the rivers \citep{Paceetal2019, Triossietal2013}, around archipelagos or islands \citep{Paceetal2012, Paceetal2019, Pulcinietal2014}, and in waters above the continental shelf and slope \citep{AZZELLINO2008}; less frequent, but still present, in deeper waters and pelagic areas. Bottlenose dolphins feed a wide range of demersal and coastal prey and can forage opportunistically behind trawling vessels \citep{Paceetal2012}. The striped dolphin is considered “pelagic” in the Mediterranean Sea, showing a general preference for highly productive, open waters beyond the continental shelf \citep{Aguilar2012}. Although the species is the most abundant cetacean in the Mediterranean, it is not found at uniform densities. The striped dolphin diet is mainly composed by pelagic or bathypelagic schooling-nictemeral fish, squids, and even crustaceans \citep{meissner2012}. 
There are not exact estimations of the number of bottlenose and striped dolphins living in the Mediterranean Sea. The poor understanding of the status of a population, together with the suspected decline in numbers (both species are listed under the status “vulnerable” in the IUCN Red List as their populations have been decreasing during the last decades), emphasize the importance of integrating all available information \citep{Paceetal2014,Paceetal2021}.

\subsection{Study area}

\begin{figure}[ht]
 \centering
    \subfloat[]{
        \includegraphics[width=0.45\textwidth]{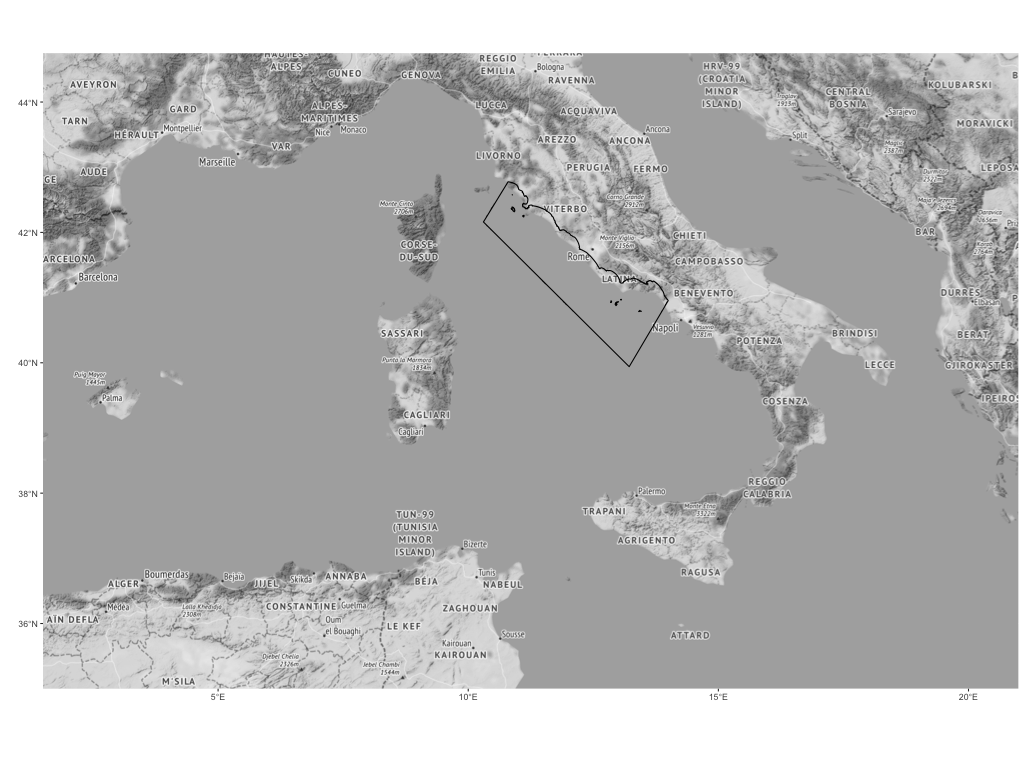}
    }%
    \qquad
    \subfloat[]{
        \includegraphics[width=0.45\textwidth]{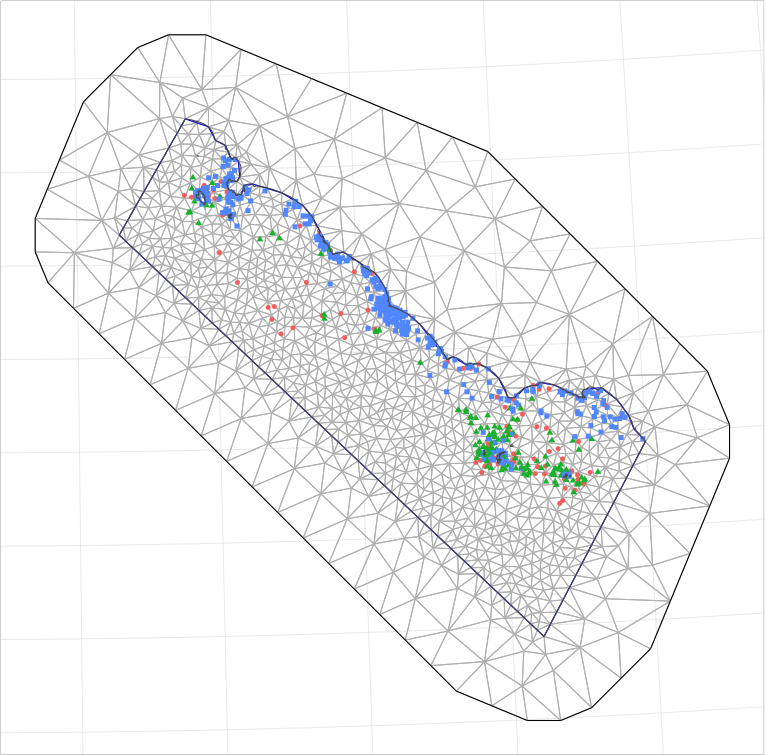}
    }%
    \caption{(a) Study area. (b) Study area and SM records for striped (green triangles), bottlenose (blue squares) dolphins and other cetacean species (red dots) superimposed to the mesh chosen for model’s estimation. }
	\label{fig:area_interest}
\end{figure}

The study area covers about 39,000 km$^2$, and is located in the central Tyrrhenian Sea (Italy) (Figure~\ref{fig:area_interest}); it is characterized by different environmental features (e.g. bathymetries), structures (e.g. seamounts) and types of habitats \citep{Paceetal2019}. Several rivers flow in this region, including the Tiber, and the simultaneous presence of both fresh and salt waters, as well as the geomorphological action of sedimentation and erosion, generate different ecological gradients, making the coastal area highly productive and rich in biodiversity \citep{Ventura2015, Ardizzone2018, Casoli2019}. The study area also includes five islands (Giglio and Giannutri at north; Ponza, Ventotene and Santo Stefano at south) and several commercial/touristic harbours generating high-levels of maritime traffic by different vessels. The region hosts seven of the eight cetacean species regularly found in the Mediterranean, with a major presence of bottlenose and striped dolphins \citep{Paceetal2019}.\\

\subsection{Data sources and attributes}

Dolphin data has been collected over 13 years (2007–2019) by three sources: a) conventional research protocols from motor and sailing boats (non‐systematic ‘haphazard’, sensu \citep{corkeron2011spatial}) (labelled UNIRM) \citep{Paceetal2019}; b) standardized monitoring protocols from platforms of opportunity within the project “FLT Mediterranean Monitoring Network” (labelled FERRY)  \citep{FLTNET,ISPRA2016, Paceetal2019, Arcangeli2019FLT}; c) social media reports (Facebook and YouTube) by sea-users \citep{Paceetal2019} (labelled SM). Data collection procedures and selection are provided in \citet{Paceetal2019}. As the SM dataset included also details on other cetacean species than the two here investigated (Figure~\ref{fig:area_interest}b ), we used this information as a proxy to infer boat densities potentially able to record the animals' presence.\\
These three sources accounted for 283 records of striped dolphin (about 50\% from SM) and 579 of bottlenose dolphin (about 80\% from SM). The major contribution by SM justified the need for a careful choice of the related model's elements. \\ 
We used distance from the coast (i.e. the euclidean distance between a sighting point and the shoreline), temperature, primary productivity, slope and depth as covariates. They are commonly selected in cetacean distribution studies as they may represent good proxies for species' ecological needs \citep{chavez2019environmental,  Stephenson2020}. Temperature and primary productivity were retrieved from COPERNICUS platform \url{https://marine.copernicus.eu/}. Depth data were downloaded from GEBCO (General bathymetric Chart of the Ocean - \url{https://www.gebco.net/}). Slope was computed from depth data through the terrain function in R. Details of the retrieved datasets and data handling procedures are reported in the Supplementary materials.

\subsection{Modelling approach}
To integrate data from all available sources and manage possible detection bias in each dataset, we adopted a point processes modelling. We followed \citet{yuan2017}, and \citet{Sicachaparada2020}, expanding their approaches by building  a Spatial Log-Gaussian Cox Process (LGCP) \citep{HandbookEco2019} that incorporates different detection functions and thinning for each data source. We assumed that sighting patterns, i.e. locations of dolphin groups in space  ($s\in {\cal S}\subset \Re^2$), are properly described by a point process whose intensity function $\lambda(s)$ is additive on the log-scale:
\begin{equation}\label{eq:logint}
    log(\lambda(s))=\mathbf{X}^T(s)\boldsymbol{\beta}+ f(\mathbf{z}) + \omega(s)
\end{equation}
where $\mathbf{X}(s)$ is a set of spatially referenced covariates, $f(\mathbf{z})$ is a smooth term (that may be present or not) of some geo-referenced covariates $\mathbf{z}$. As prior for $f(\mathbf{z})$ a common choice is a random walk of order 1 \citep{RueHeld:2005}. Finally,  $\omega(s)$ is a zero-mean Gaussian process describing the residual spatial variation. As in \citet{yuan2017} we adopted a Mat\`ern covariance of order 1 with range $\rho$ and standard deviation $\sigma$. 

We assume that the above process is observed in three different ways, conditionally independent given $\lambda(s)$. Thus, 3 observed intensities were defined:
\begin{equation}\label{eq:obs_models}
    \lambda^*(s)=t_jg_j(s)\lambda(s),\qquad j=1,2,3
\end{equation}
where $t_j$ is a time scaling factor and $g_j(s)$ is the detection function (with values between $0$ and $1$)  which determines the original process's thinning. The form of the detection function depends on the type of observational process. For the UNIRM data, we set
\begin{eqnarray}
g_1(s)&=&\left\{\begin{array}{cc}
    1 & d_1(s)\le K \\
    0 &  d_1(s)>K
\end{array}\right.\label{eq:detRM}\\
\end{eqnarray}
where $d_1(s)$ is the distance (Km) between point $s$ and the position of the boat when the group was sighted. $K$ was defined as the maximum distance measured between the location of the first visual sight of a dolphin group by researchers (equipped with 7x50 and 10x50 binoculars) on the boat and the effective location of the group under optimal survey conditions (i.e., sea state$\le$1 Douglas, wind force$\le$1 Beaufort, no rain, no fog, no clouds). This measurement was possible because, upon sighting dolphins, researchers marked the GPS point where the animals were first located, the survey effort was suspended, and the vessel departed from its route to approach the group to a suitable distance (10-30 m) to correctly identify the species, estimate group size and composition.  $K$ was set to 4 Km, assuming that researchers can spot animals closer than $K$. \\
For the FERRY data, we used the classical half normal detection function \citep{Thompson1987}  defined as 
\begin{eqnarray}
g_{2}(s)&=&\exp\left(-\frac{d_2(s)^2}{2\ \xi_2^2}\right) \label{eq:detferry}
\end{eqnarray}
where, $d_2(s)$  is the perpendicular distance (Km) to the ferry track and $\xi$ is a scale parameter. \\

\begin{figure}[ht]
  \subfloat[]{
	\begin{minipage}[c][1\width]{
	   0.3\textwidth}
	   \centering
	   \includegraphics[width=1\textwidth]{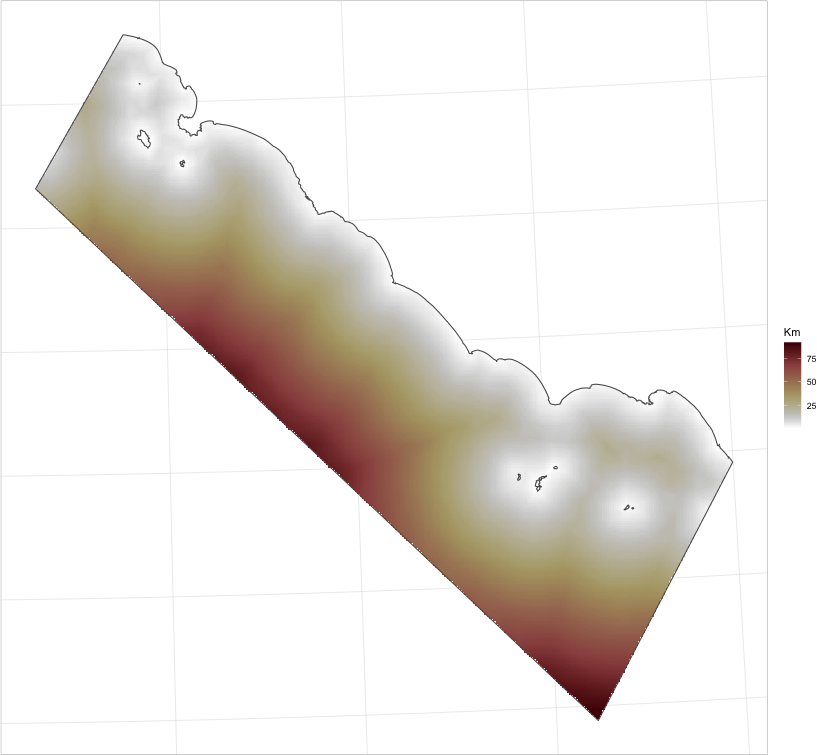}
	\end{minipage}}
 \hfill	
  \subfloat[]{
	\begin{minipage}[c][1\width]{
	   0.3\textwidth}
	   \centering
	   \includegraphics[width=1\textwidth]{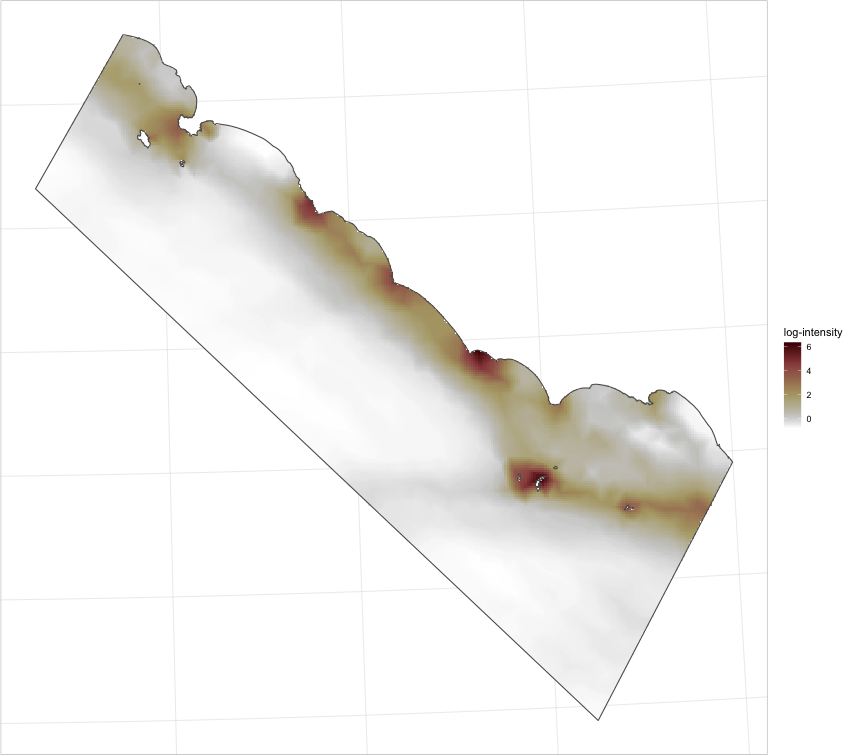}
	\end{minipage}}
	 \hfill 	
  \subfloat[]{
	\begin{minipage}[c][1\width]{
	   0.3\textwidth}
	   \centering
	   \includegraphics[width=1\textwidth]{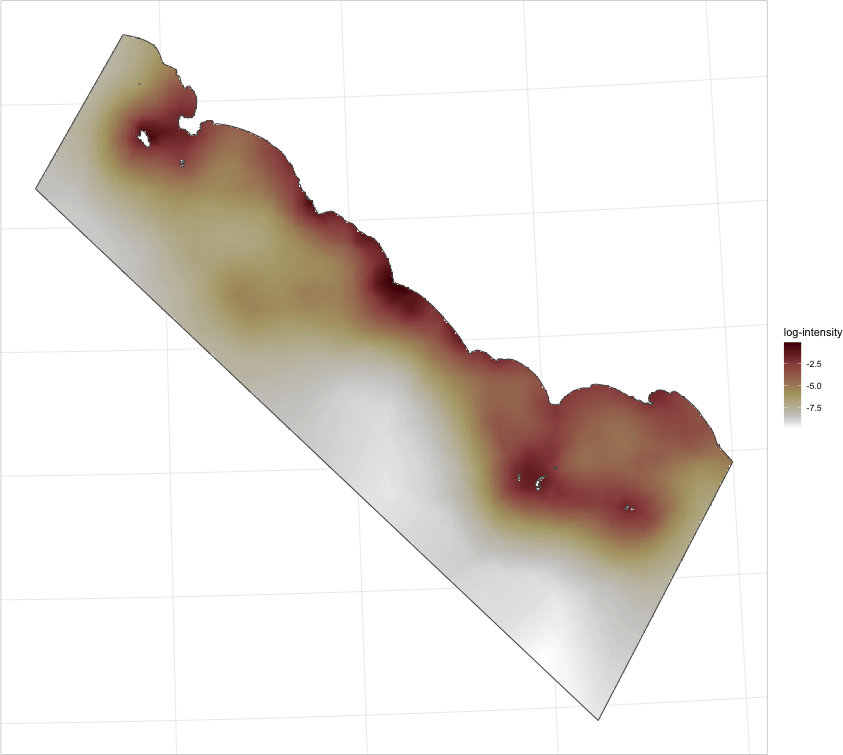}
	\end{minipage}}
\caption{(a) distance from the coastline, (b) log vessels density, (c) estimated log- intensity from observations of all species}
\label{fig:int_surfaces}
\end{figure}

For the SM dataset, the definition of the detection function has been carefully considered for biases. Records in this dataset are affected by large uncertainty, as observations are generally (a) skewed towards more accessible areas  \citep{Sicachaparada2020, monserrat2019} and (b) collected from small leisure boats that are difficult to track in a systematic way. To better define “more accessible” and consider the distribution of the small boats we explored three different possibilities. \\
First, we reasonably assumed that locations closer to the coast are more accessible to sea-users with small boats. Thus, following \citep{Sicachaparada2020}, the detection function was defined as:
\begin{equation}\label{eq:detdistcoast}
    g_{3,1}(s) = \exp\left(-\frac{d_{3,1}^2(s)}{2\ \xi_{3,1}^2}\right)
\end{equation} 
where $d_{3,1}(s)$ is the distance from the coast (Figure~\ref{fig:int_surfaces}a) and $\xi_{3,1}$ a scaling parameter.
However, the distance from the coast may not provide an accurate representation of the small boats' density in a given area: locations close to harbours and holiday destinations (e.g. islands) are generally more crowded than other sites at the same distance from the coastline.\\
Secondly, to obtain information on the boats’ density in the study area, we used data from EMODnet  (European Marine Observation and Data Network; \citep{EMODNET2019}, a free-usage platform of vessel density data derived from boats using AIS (Automatic Identification System, mandatory above 15m length). The database has a spatial resolution of 11 Km and covers 2017-2019 period. We selected 2 vessels’ categories (sailboats and pleasure crafts) from the 11 listed, and applied a kernel estimator to ensure a smoothed density surface. The resulting log-density surface (Figure~\ref{fig:int_surfaces}b) was labelled as “vessel log-density surface”. The detection function was defined as

\begin{equation}\label{eq:detvessel}
    g_{3,2}(s) = \Phi(\frac{d_{3,2}(s)}{\xi_{3,2}} -\mu_{3,2})
\end{equation}
where $d_{3,2}(s)$ is the log of estimated boats density, and $\Phi$ is the normal cumulative distribution function (cdf) with $\mu_{3,2}$ and $\xi_{3,2}$ as location and scale parameters, respectively. The normal cdf was selected as we required the detection function to be close to 1 when the vessel log-density is high, and to (or equal to) zero when it is small (or null). As expected, higher vessel (log) densities were identified near the principal harbours and the islands (Figure~\ref{fig:int_surfaces}b).  However, EMODnet information accounted for a limited timeframe compared to our study and for larger vessels than the ones generally reporting observation records in SM platforms (small recreational boats moving near the coastline). \\
Lastly, we used the entire SM dataset of 581 records (125 striped and  334 bottlenose dolphins and 122 other cetacean species) to estimate the observation process’s intensity. We considered the spatial pattern of such observations as a proxy for the small boat density process if we disregard the species. A similar approach was used in occupancy models context, where non-detection records were constructed from sightings of other ”benchmark” species (see for example \citet{Dennis_et_al2017, Kery_et_al2010}).
We applied a spatial LGCP to estimate the (log) intensity of the process. Details of the estimation process can be found in the  Supplementary material. Figure \ref{fig:int_surfaces}c shows the resulting estimated log-intensity that we then used as input for the detection function:
\begin{equation}\label{eq:detanimals}
    g_{3,3}(s) = \Phi(\frac{d_{3,3}(s)}{\xi_{3,3}} -\mu_{3,3})
\end{equation}
where  $d_{3,3}(s)$ is the estimated log-intensity at point $s$ while $\Phi$, $\mu_{3,3}$ and $\xi_{3,3}$ are defined as in \eqref{eq:detvessel}. \\
Eventually, another potential bias affecting the observation processes is the different time (days) spent at sea by each data source. To account for this we introduced the  $t_j$ parameter in expression~\eqref{eq:logint}. $t_j$ is known for both the FERRY and the UNIRM data, but it is indeterminate for SM data. We know that SM observations were collected by leisure boats all over the year, with a major number of sightings reported in spring-summer. Thus,  we ran estimations with $t_3=160, 200, 365$ days, without sensible changes, and selected $t_3=360$.

\subsection{Priors specification}
To finalize the model in a Bayesian framework, we needed to specify priors for all model parameters. To avoid identifiability issues when estimating both the animal intensity and the observation process, we used slightly informative priors.
For the parameters in the spatial field $\omega(s)$ in \eqref{eq:logint} we used PC priors \citep{Fuglstad_etal2019} setting $P(\rho<150)=0.5$ and $P(\sigma>2) = 0.01$. Meaning  that we consider  a standard deviation above 2 as large and a range of 150 Km likely. 
We assing $\mathbf{\beta}$ Gaussian prior with mean 0 and precision 0.01. The location parameters $\mu$ in \eqref{eq:detvessel} and \eqref{eq:detanimals} are also Gaussian  with mean 0 and precision 0.01. Finally, for the scale parameters in \eqref{eq:detferry}-\eqref{eq:detvessel},  let $\xi = F^{-1}_{\alpha}(\Phi(\theta)$ where $F^{-1}(\cdot)$ is the inverse exponential cdf with rate $\alpha$ and
$\Phi$ is a normal cdf. This corresponds to assigning an exponential prior to $\xi$. We then assign $\theta$ a standard normal prior. The parameter $\alpha$ is set to $1/20$ in \eqref{eq:detdistcoast}, and $1$ in all other cases. The difference in rate is due to the different scale of the three inputs for the detection function (Figure~\ref{fig:int_surfaces}).

\subsection{Inference and computational approach}

The traditional way of fitting point processes is by gridding the space and modelling the intensity on a  discrete number of cells. This implies that observations' locations are also approximated. We followed instead the approach introduced in \citet{simpsonetal2016} and applied in \citet{yuan2017}  and \citet{Sicachaparada2020}. Such an approach allowed us to use the true sighting locations, thus avoiding loss of information. Besides,  the Gaussian field's SPDE representation has several computational advantages (see \citet{lindgrenSPDE}). To build a spatial model using the SPDE approach, we used the mesh shown in Figure \ref{fig:area_interest}b.  

For computational efficiency, we used the INLA method for numerical Bayesian inference with Gaussian Markov random fields \citep{rueetal2009}. INLA allowes also to easily combine the three observation model in \ref{eq:obs_models} to form the likelihood. Our model does not directly fall under the latent Gaussian model framework for the INLA estimation software because the parameters in the detection functions in \eqref{eq:detferry}-\eqref{eq:detvessel} do not enter the model in a log-linear way. We use therefore the methodology introduced in \citet{yuan2017} and implemented in the {\tt inlabru} R package \citep{inlabru_paper} that allows fitting models with some non-linear elements. This is done by linearizing the model via Taylor approximation and using a line search to optimize the linearization point.\\
Model evaluation was carried out using goodness of fit measures as in \citet{Sicachaparada2020}, through the Deviance Information Criterion (DIC), Watanabe-AiKaike Information Criterion (WAIC), Marginal Likelihood (MLIK), and the logarithm of the pseudo marginal likelihood (LMPL). As a benchmark for the SM detection function choice, we use a constant detection function $g(s)=1,\;\forall s$, that is equivalent to not include any thinning for the SM data.

\section{Results}\label{sec:Results}

\begin{figure}[ht]
    \centering
	\includegraphics[width=.7\textwidth]{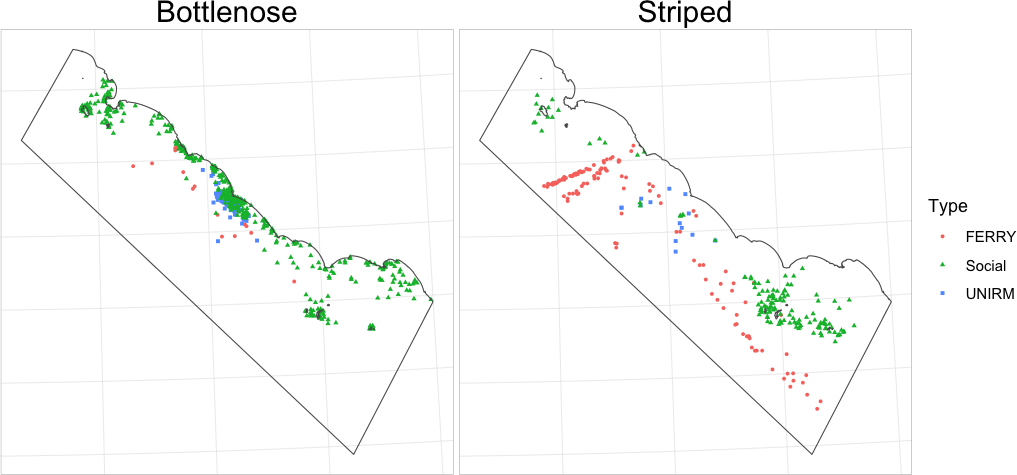}
    \caption{Dolphins encounters' locations by source (observation processes): SM (green), FERRY (red), and UNIRM (blue).}
    \label{fig:observations}
\end{figure}

The distribution of the dolphins encounters in the study area is shown in Figure~\ref{fig:observations}. Environmental covariates’ selection was finalized considering several combinations of detection functions. Two different models have been selected, one for each species 

\begin{itemize}
    \item[] \emph{S. coeruleoalba} (Striped dolphin)
    \begin{itemize}
        \item Depth: categorized as  ($<100m$, $100m - 200m$, $200m-1000m$, $>1000m$)
        \item Slope: non parametric with a prior Random walk of order 1
        \item Distance from the coast: linear term
    \end{itemize}
    \item[] \emph{T. truncatus} (Bottlenose dolphin)
    \begin{itemize}
        \item Depth: linear term
        \item Slope: linear term
        \item Distance from the coast: linear term
    \end{itemize}
\end{itemize}

%
%

\begin{figure}%
    \centering
    \subfloat[]{
        \includegraphics[width=0.45\textwidth]{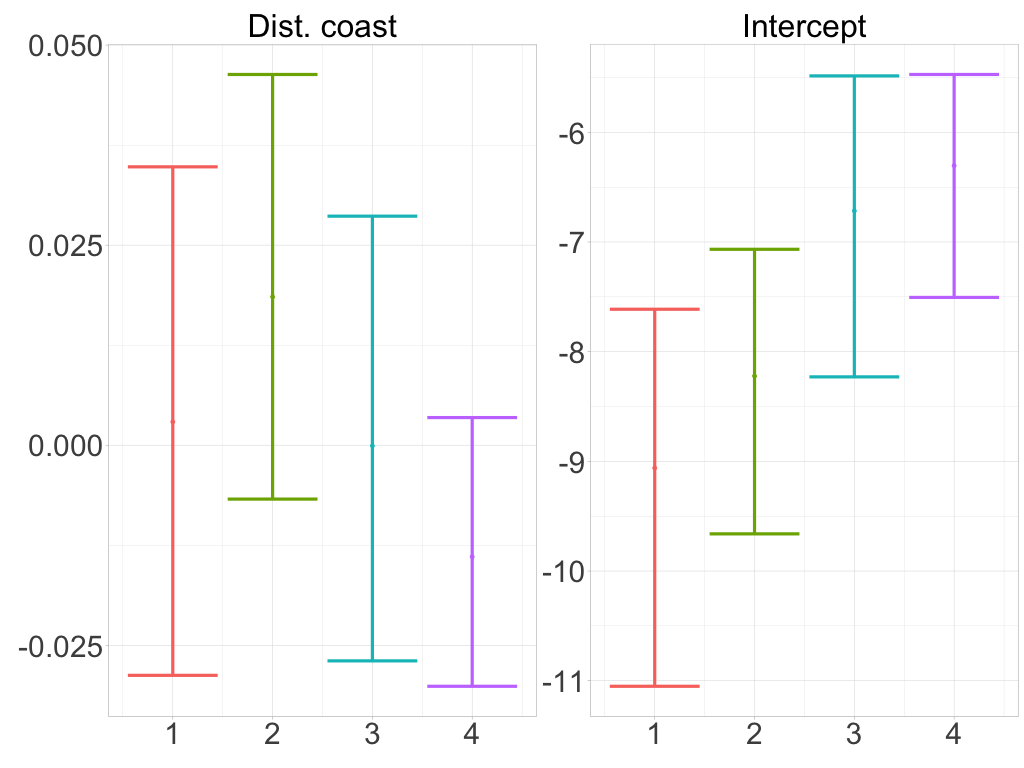}
    }%
    \qquad
    \subfloat[]{
        \includegraphics[width=0.45\textwidth]{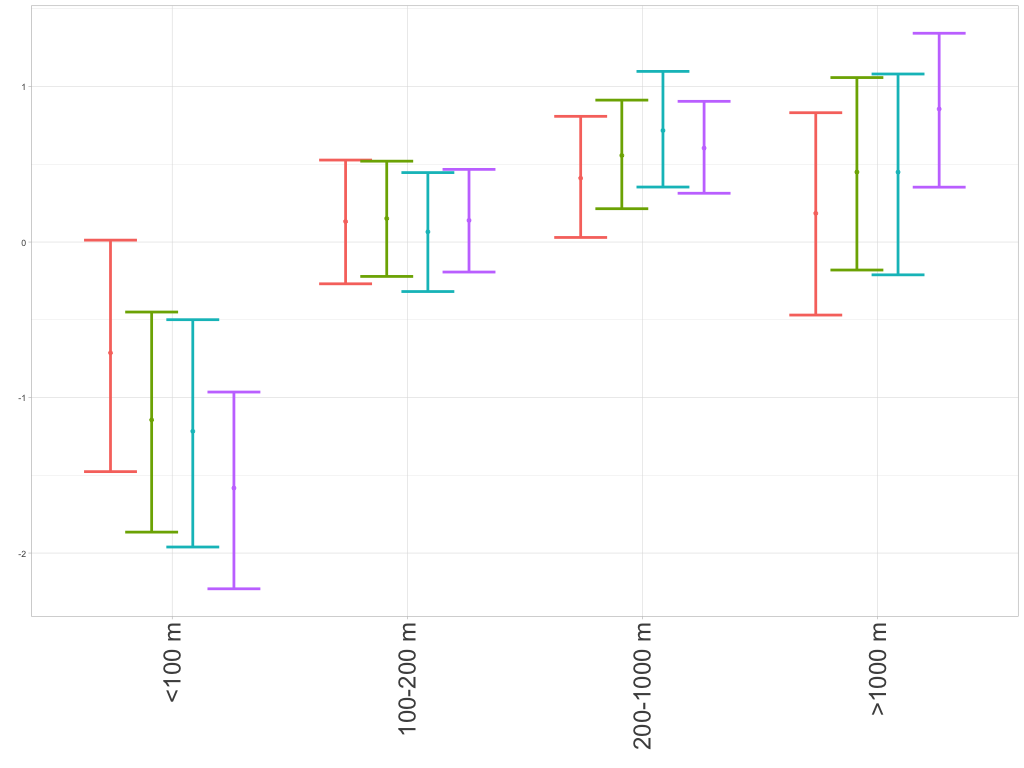}
    }%
    \qquad
    \subfloat[]{
        \includegraphics[width=0.45\textwidth]{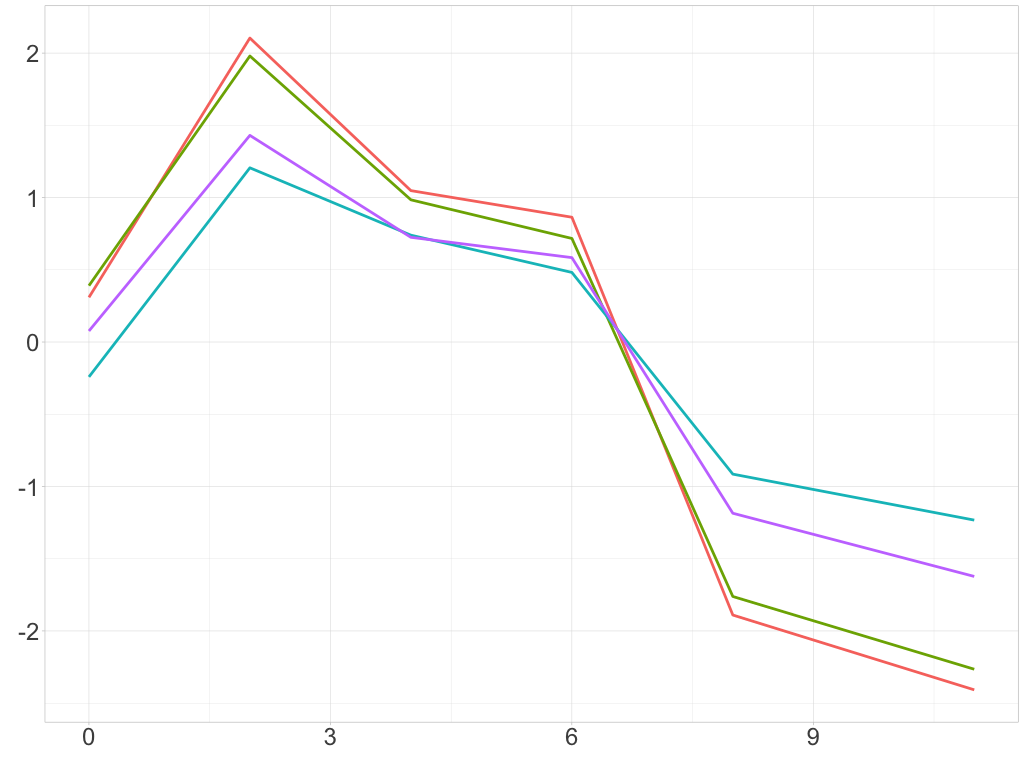}
    }%
    \caption{Estimated intercept and effect of the distance from the coastline (a) effect of depth (b) and slope (c) for the four fitted models for the striped  dolphins. In (a) and (b) the error bars indicate 95\% credible intervals using different detection functions for the SM data, detection constant (red), detection \eqref{eq:detdistcoast} (green),  detection \eqref{eq:detanimals} (blue) and  detection \eqref{eq:detvessel} (violet). In (c) the estimated smooth terms for the slope are reported}\label{fig:stenellaModelEffects}%
\end{figure}

The evaluation of SM detection functions is based on model's goodness of fit measures, DIC WAIC, MLIK and the LMPL, it is reported in table~\ref{tab:detectionchoice}. The selected best performing detection function for all criteria and species is \eqref{eq:detanimals} (labelled as intensity). This choice affected model’s terms estimate.  
For striped dolphin model with varying detection functions (Fig.~\ref{fig:stenellaModelEffects}), the effects of categorized Depth were in agreement with the species’ distribution ranges: it is generally not found in very shallow waters (negative effects), observed at 100-200m depth, and more often encountered at depths over 200m. The effect of the detection function was found in the smaller size of the credible intervals with detection \eqref{eq:detanimals}. Slope showed a significant reduction effect in the encounters where it is steeper. No significant difference was found among the smooth effects with varying detection (overlapping 95\% confidence band, not shown). The effect of the Distance from the coast was not significant, and the intercept was larger for detection functions  \eqref{eq:detvessel} and \eqref{eq:detanimals}, with the latter showing less uncertainty than the first.

\begin{figure}%
  \includegraphics[width=0.9\textwidth]{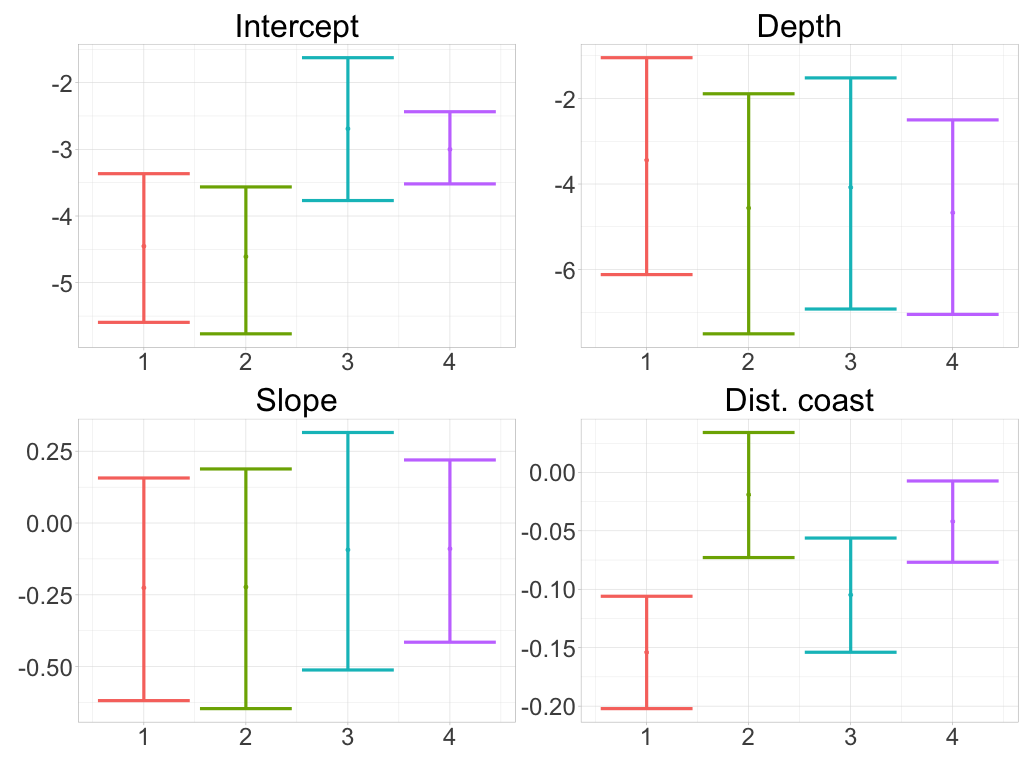}
    \caption{Estimated intercept and covariate effects for the four fitted models for the bottlenose dolphin. The error bars indicate 95\% credible intervals using different detection functions for the SM data, detection constant (red),detection \eqref{eq:detdistcoast} (green),  detection  \eqref{eq:detanimals} (blue) and  detection \eqref{eq:detvessel} (violet).} 
    \label{fig:tursiope_effect}%
\end{figure}

For bottlenose dolphin model with varying detection functions (Fig.~\ref{fig:tursiope_effect}), both Depth and Distance from the coast had negative effects on sightings (deeper waters and increasing distance from the coast mean less encounters), with no-significant difference among detection functions. Again, detection \eqref{eq:detanimals} induced narrower 95\% credible intervals.
Estimates of detection functions’ parameters are reported in tables~\ref{tab:StenellaModel} and \ref{tab:TursiopeModel} in the Supplementary Materials. 

As a measure of relative uncertainty for the predicted intensity $\lambda(s)$ we use the relative width of the 50\% posterior credible interval (RWPCI) as proposed in \cite{yuan2017}. This measure is defined as the interquartile range divided by the median
\begin{equation}\label{eq:RWPCI}
\text{RWPCI} = (\text{Q}_3-\text{Q}_1)/\text{Q}_2.
\end{equation}

\begin{figure}%
  \includegraphics[width=0.9\textwidth]{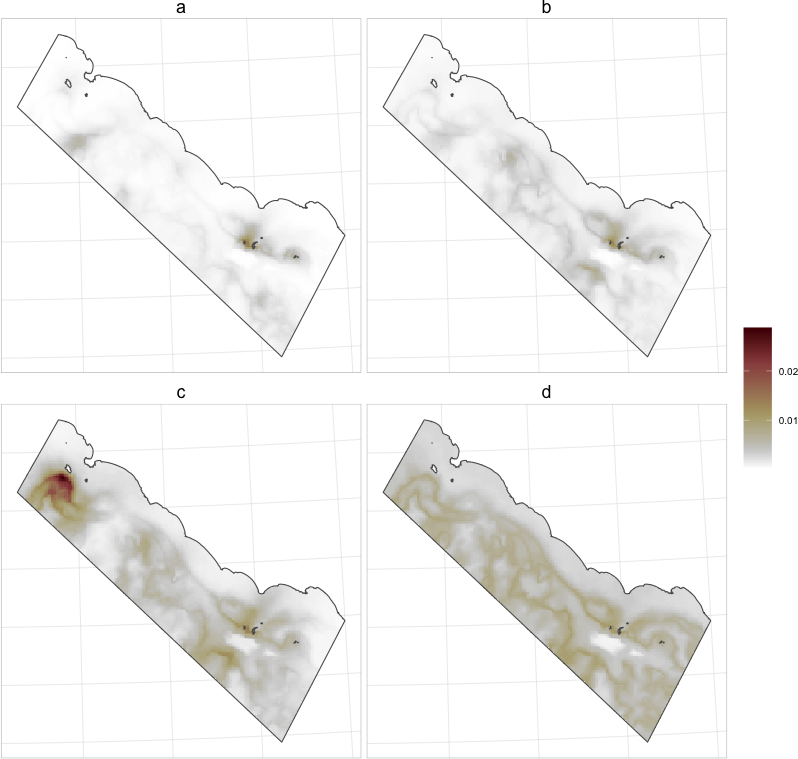}
    \caption{Estimated posterior median for the intensity of striped dolphins using different detection functions for SM data. (a) constant detection, (b) detection  \eqref{eq:detdistcoast},  (c) detection  \eqref{eq:detvessel}, (d) detection \eqref{eq:detanimals}. } \label{fig:stenella_int_surface}%
\end{figure}

\begin{figure}%
  \includegraphics[width=0.9\textwidth]{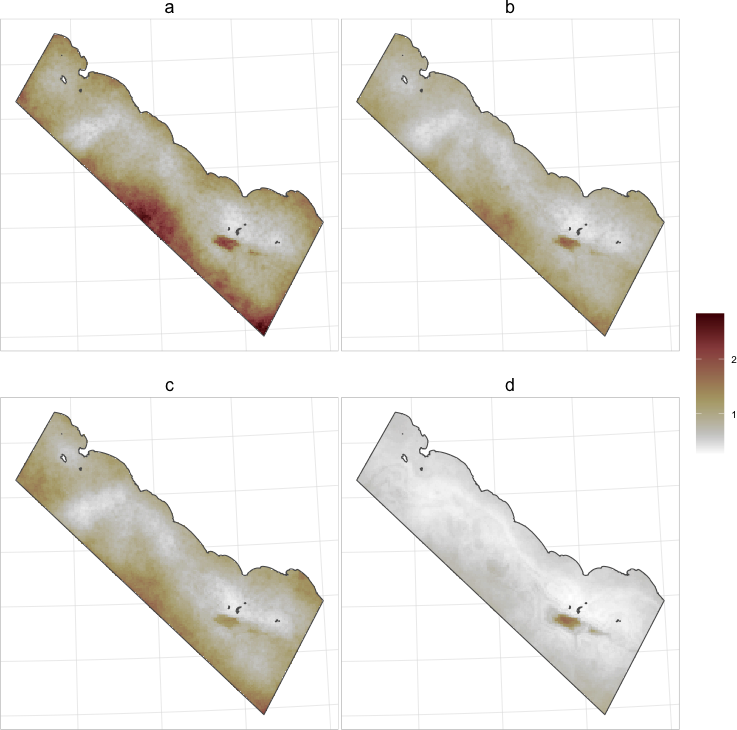}
    \caption{RWPCI for the intensity of striped dolphin using different detection functions for SM data. (a) constant detection, (b) detection  \eqref{eq:detdistcoast},  (c) detection  \eqref{eq:detvessel}, (c) detection  \eqref{eq:detanimals}. } \label{fig:stenella_cv_surface}%
\end{figure}

\begin{figure}%
  \includegraphics[width=0.9\textwidth]{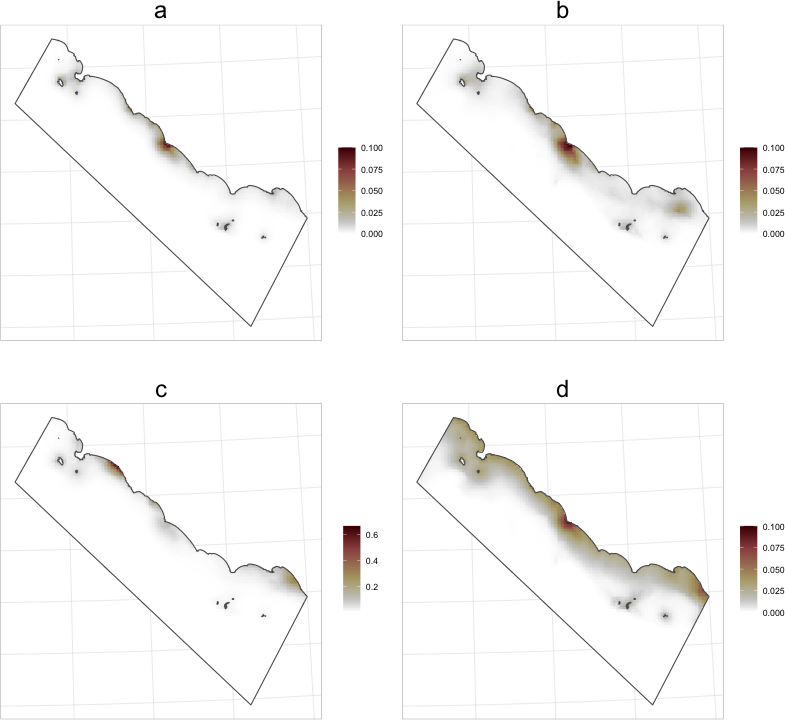}
    \caption{Estimated posterior median for the intensity of bottlenose dolphin using different detection functions for SM data. (a) constant detection, (b) detection  \eqref{eq:detdistcoast},  (c) detection  \eqref{eq:detvessel}, (c) detection \eqref{eq:detanimals}. Note that the scale in (c) is different from the other three figures.}%
    \label{fig:tursiope_int_surface}%
\end{figure}

\begin{figure}%
  \includegraphics[width=0.9\textwidth]{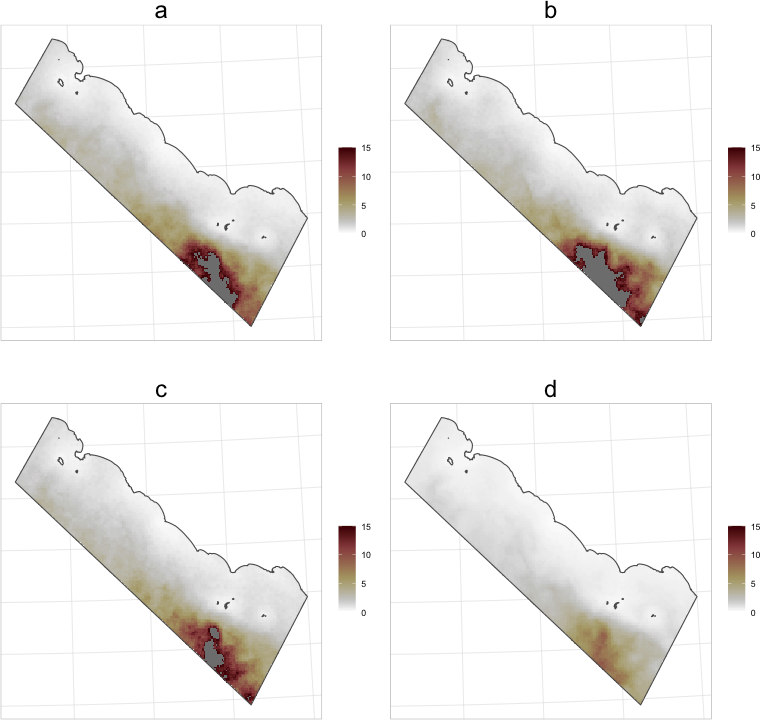}
    \caption{RWPCI for the intensity of bottlenose dolphin using different detection functions for SM data. (a) constant detection, (b) detection  \eqref{eq:detdistcoast},  (c) detection \eqref{eq:detvessel}, (c) detection \eqref{eq:detanimals}. The colour palette is cut off at 15 to exclude the extreme values in the bottom right corner. }%
    \label{fig:tursiope_cv_surface}%
\end{figure}

\begin{figure}
    \centering
    \includegraphics[width=0.8\textwidth]{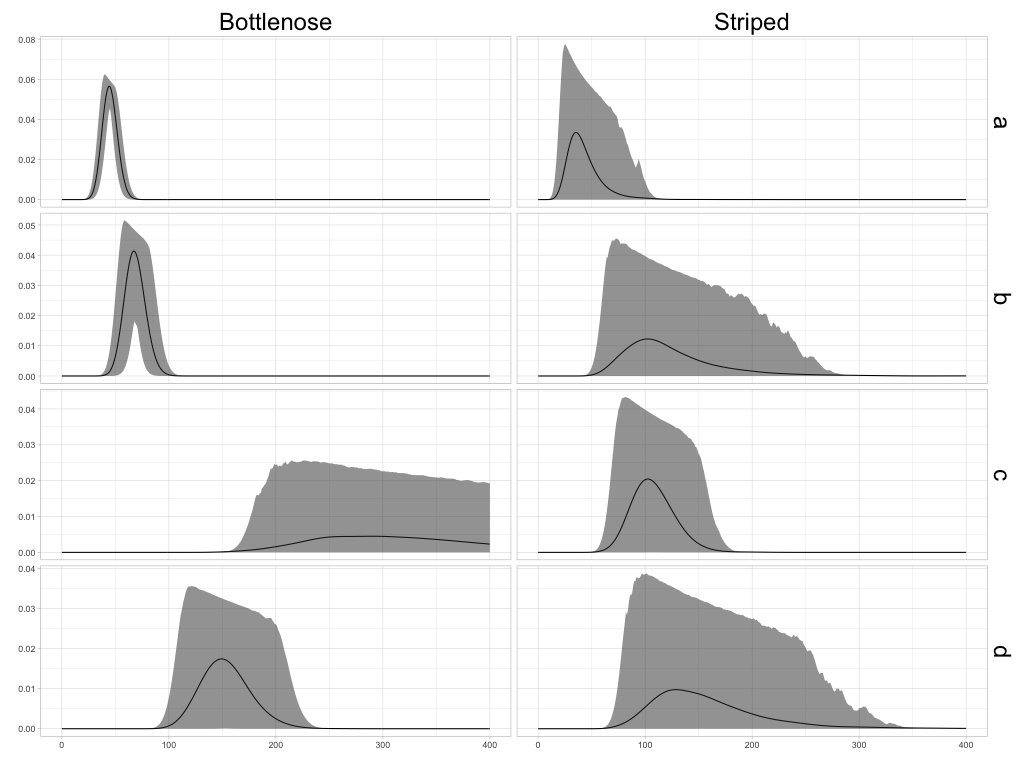}
    \caption{Estimated probabilities for the number of sightings over one year in the study area for bottlenose  (left) and striped (right) dolphins for the four fitted models: (a)  constant detection, (b) detection  \eqref{eq:detdistcoast} ,(c) detection \eqref{eq:detvessel} and (d) detection  \eqref{eq:detanimals}. The grey band indicate 97\% credible intervals. }
    \label{fig:Nanimals}
\end{figure}

The intensity surfaces estimated for the striped and bottlenose dolphins are shown in Figures  \ref{fig:stenella_int_surface} and \ref{fig:tursiope_int_surface}, respectively; associated RWPCIs \eqref{eq:RWPCI} are mapped in Figures  \ref{fig:stenella_cv_surface} and \ref{fig:tursiope_cv_surface}. The intensity surface for both species changed consistently with the different detection function adopted for SM. For example, the vessel-based detection \eqref{eq:detvessel} (vessel-log density surface) induced some artifacts for the striped dolphin, and in general over-estimated the dolphins’ encounter probability, while the detection based on distance from the coast \eqref{eq:detdistcoast} and the constant detection, under-estimate the same probability for the striped dolphin and create some artifacts. A relevant feature of the detection function \eqref{eq:detanimals} is that allows a consistent reduction in the uncertainty associated to the estimated intensity surface.\\ In Fig. \ref{fig:Nanimals} we describe the estimated probabilities for the number of sightings over one year for both species. Panel (d) corresponds to the chosen detection, and is showing an ecologically sound distributions for the number of sightings. Striped dolphins are more common in the area than the bottlenose dolphin. The smaller number of sightings in the database (283 striped and 579 bottlenose dolphins) induces a large uncertainty on the estimates, however it still allows for correctly capturing the species spatial distribution.

\begin{table}[ht]
\centering
\begin{tabular}{lrrrr}
  \hline
  model & DIC & WAIC & MLIK & LMPL \\ 
  \hline
  \multicolumn{5}{c}{(a) Stenella coeruleoalba}\\\hline
  
  constant detection & 4078.53 & 4129.16 & -2111.81 & -2098.13 \\ 
 detection \eqref{eq:detdistcoast} & 3895.52 & 3933.01 & -2008.44 & -1988.77 \\ 
 detection \eqref{eq:detvessel} & 3840.93 & 3889.51 & -2019.20 & -1969.47 \\ 
  detection \eqref{eq:detanimals} & 3789.38 & 3810.08 & -1942.07 & -1922.56 \\ 
    \hline
   \multicolumn{5}{c}{(b) Tursiops truncatus}\\
   \hline
    constant detection & 4639.94 & 4874.47 & -2375.33 & -2568.07 \\ 
   detection \eqref{eq:detdistcoast} & 4555.23 & 4797.14 & -2337.60 & -2726.91 \\ 
   detection \eqref{eq:detvessel}& 4552.61 & 4810.37 & -2344.68 & -2658.98 \\ 
   detection \eqref{eq:detanimals} & 4485.78 & 4624.58 & -2281.27 & -2351.19 \\ 
 
   \hline
\end{tabular}
\caption{Comparison criteria for the four fitted model for both striped  (a)  and bottlenose (b) dolphins.}\label{tab:detectionchoice}
\end{table}

\section{Discussion}\label{sec:discussion}
This study demonstrates that methods of spatial data integration able to carefully consider and minimize datasets' biases can be efficiently used to predict species' distribution. Results here obtained may be broadly applicable to other species that require an improvement of spatial knowledge for their conservation and management. \\
\citet{Dorazio2014} pointed out that several statistical models have been proposed to analyse presence-only data, but they have largely ignored the effects of imperfect detectability and survey bias. The same author showed that proper modelling choices could reduce the bias in SDM estimates induced by these types of errors. Here we do more than just correct for detectability issues; we allow multiple sources of information to be integrated. We defined and estimated source-specific detection functions considering the nature of the data, i.e. presence-only, and the different observation processes, offering a more precise picture of the distribution of two dolphin species in the central Mediterranean. The output is consistent with the ecology of these species, highly supporting a thoughtful usage of spatial data extracted from social media platforms and introducing a novel way to model observation biases.
In analysing different detection functions, we optimise distribution models for each species. That is very attractive considering the importance of defining suitable habitats for vulnerable or endangered cetaceans exposed to anthropogenic disturbance or threats, particularly in coastal areas \citet{pace2018}. 

The point process approach allows us to reliably estimate the observation intensity surface. The analysis of intensity surfaces in figures \ref{fig:stenella_int_surface} and \ref{fig:tursiope_int_surface}, gives important insights on the relevance of the detection function in observation intensity estimation. The artefacts around the Tiber river estuary (central part of the area) for the bottlenose dolphin and close to the Giglio island (northern portion of the study area) for the striped dolphin are solved by detection \eqref{eq:detanimals}. Again, with the same detection function's choice, analysing figures \ref{fig:stenella_cv_surface} and \ref{fig:tursiope_cv_surface}, we can observe the reduction of intensity estimates' variability (and hence uncertainty). The proposed "best'' choice is very general and can be adopted whenever social media data are available. \\
The two species were also studied in \citet{Paceetal2019} using a presence-only data approach based on MaxEnt \citep{maxent2006}. While results related to the bottlenose dolphin analysis were ecologically sound and coherent, striped dolphins analysis was unfeasible in that framework, given the relevant number of near-to-the-coast observation by sea-users. In particular, the depth around the Pontine islands rapidly increases with the distance from the coast, playing a misleading role in the MaxEnt modelling approach. The proposed methodology, instead, is fully able of capturing both species behaviour, thus addressing the complex task of finding targeted techniques weighting species' diversity.  \\
Some limitations are intrinsic to the proposed approach. On the one hand, spatial estimation does not distinguish between land and sea. The latter implies the use of post-processing to cut the estimated intensity surface. On the other, each analysed detection function is not very flexible. Eventually, the information used to model the observation effort in the SM data can be further improved.
Hence, further investigations will be carried out to:
\begin{itemize}
    \item Develop spatially non-stationary modelling approaches where a barrier can be added at the coastline as in \citet{BAKKA2019}.
    \item Develop flexible detection functions
    \item Explore the use of satellite data to estimate the density of small boats in the study area \citep{remotesensing2017}
\end{itemize}

The implementation of these tasks and the improvement of the models’ capabilities may further develop a fast-growing research approach and provide innovative insights in marine top-predators’ distribution patterns. The multiplicity of issues confronting these marine species requires collaborative efforts at all levels to share and merge resources, data, and expertise efficiently  \citep{Vella2021common, pace2018}.



\section*{Acknowledgements}
This work was partially supported by project \emph{Joint Cetacean Database and Mapping (JCDM) in Italian waters: a tool for knowledge and conservation}, Sapienza University of Rome (nr. RM1201729F23D51B).
\section*{Data Accessibility Statement}
The data presented in this study are fully available to any qualified researcher on request from the corresponding author.

\section*{Conflict of interest}
The authors declare no conflict of interest
\bibliographystyle{plainnat}
\bibliography{main}

\begin{thebibliography}{58}
\providecommand{\natexlab}[1]{#1}
\providecommand{\url}[1]{\texttt{#1}}
\expandafter\ifx\csname urlstyle\endcsname\relax
  \providecommand{\doi}[1]{doi: #1}\else
  \providecommand{\doi}{doi: \begingroup \urlstyle{rm}\Url}\fi

\bibitem[Aguilar and Gaspari(2012)]{Aguilar2012}
A.~Aguilar and S.~Gaspari.
\newblock \textit{Stenella coeruleoalba}. the iucn red list of threatened
  species 2012: e.t20731a2773889.
\newblock \url{https://www.iucnredlist.org/species/20731/50374282}, 2012.
\newblock Accessed: 2020-12.

\bibitem[Arcangeli et~al.(2019)Arcangeli, Aissi, Atzori, Azzolin, Campana,
  Carosso, Crosti, David, Di~Meglio, Frau, Garcia~Garin, Giacoma, Paraboschi,
  Pellegrino, Rosso, Roul, Sara, Scuderi, Tepsich, Tringali, and
  Vighi]{Arcangeli2019FLT}
A.~Arcangeli, M.~Aissi, F.~Atzori, M.~Azzolin, I.~Campana, L.~Carosso,
  R.~Crosti, L.~David, N.~Di~Meglio, F.~Frau, O.~Garcia~Garin, C.~Giacoma,
  M.~Paraboschi, G.~Pellegrino, M.~Rosso, M.~Roul, G.~Sara, A.~Scuderi,
  P.~Tepsich, M.~Tringali, and M.~Vighi.
\newblock Fixed line transect mediterranean monitoring network (flt med net),
  an international collaboration for long term monitoring of macro-mega fauna
  and main threats fixed line transect mediterranean monitoring network.
\newblock \emph{Biologia Marina Mediterranea}, 26\penalty0 (1):\penalty0
  400--401, 2019.

\bibitem[Ardizzone et~al.(2018)Ardizzone, Belluscio, and
  Criscoli]{Ardizzone2018}
G.D. Ardizzone, A.~Belluscio, and A.~Criscoli.
\newblock \emph{Atlante degli Habitat dei Fondali Marini del Lazio.}
\newblock Sapienza Universit{\`a} Editrice: Rome, Italy., 2018.

\bibitem[Azzellino et~al.(2008)Azzellino, Gaspari, Airoldi, and
  Nani]{AZZELLINO2008}
A.~Azzellino, S.~Gaspari, S.~Airoldi, and B.~Nani.
\newblock Habitat use and preferences of cetaceans along the continental slope
  and the adjacent pelagic waters in the western ligurian sea.
\newblock \emph{Deep Sea Research Part I: Oceanographic Research Papers},
  55\penalty0 (3):\penalty0 296 -- 323, 2008.
\newblock ISSN 0967-0637.
\newblock \doi{https://doi.org/10.1016/j.dsr.2007.11.006}.
\newblock URL
  \url{http://www.sciencedirect.com/science/article/pii/S0967063707002609}.

\bibitem[Bachl et~al.(2019)Bachl, Lindgren, Borchers, and
  Illian]{inlabru_paper}
Fabian~E. Bachl, Finn Lindgren, David~L. Borchers, and Janine~B. Illian.
\newblock inlabru: an r package for bayesian spatial modelling from ecological
  survey data.
\newblock \emph{Methods in Ecology and Evolution}, 10\penalty0 (6):\penalty0
  760--766, 2019.
\newblock \doi{https://doi.org/10.1111/2041-210X.13168}.
\newblock URL
  \url{https://besjournals.onlinelibrary.wiley.com/doi/abs/10.1111/2041-210X.13168}.

\bibitem[Bakka et~al.(2019)Bakka, Vanhatalo, Illian, Simpson, and
  Rue]{BAKKA2019}
Haakon Bakka, Jarno Vanhatalo, Janine~B. Illian, Daniel Simpson, and Håvard
  Rue.
\newblock Non-stationary gaussian models with physical barriers.
\newblock \emph{Spatial Statistics}, 29:\penalty0 268--288, 2019.
\newblock ISSN 2211-6753.
\newblock \doi{https://doi.org/10.1016/j.spasta.2019.01.002}.
\newblock URL
  \url{https://www.sciencedirect.com/science/article/pii/S221167531830099X}.

\bibitem[Bearzi et~al.(2012)Bearzi, Fortuna, and Reeves]{Bearzi2012}
G.~Bearzi, C.M Fortuna, and R.R Reeves.
\newblock \textit{Tursiops truncatus}. the iucn red list of threatened species
  2012: e.t22563a2782611.
\newblock \url{https://www.iucnredlist.org/species/22563/156932432}, 2012.
\newblock Accessed: 2020-12.

\bibitem[Bolzon et~al.(2020.)Bolzon, Cossarini, Lazzari, Salon, Teruzzi,
  Feudale, Di~Biagio, and Solidoro]{Bolzon2020}
G.~Bolzon, G.~Cossarini, P.~Lazzari, S.~Salon, A.~Teruzzi, L.~Feudale,
  V.~Di~Biagio, and C.~Solidoro.
\newblock Mediterranean sea biogeochemical analysis and forecast (cmems
  med-biogeochemistry 2018-present).
\newblock \emph{Copernicus Monitoring Environment Marine Service (CMEMS)},
  2020.
\newblock
  \doi{https://doi.org/10.25423/CMCC/MEDSEA_ANALYSIS_FORECAST_BIO_006_014_MEDBFM3.}

\bibitem[Breen et~al.(2017)Breen, Brown, Reid, and Rogan]{breen2017}
Patricia Breen, Susie Brown, David Reid, and Emer Rogan.
\newblock Where is the risk? integrating a spatial distribution model and a
  risk assessment to identify areas of cetacean interaction with fisheries in
  the northeast atlantic.
\newblock \emph{Ocean \& Coastal Management}, 136:\penalty0 148--155, 2017.

\bibitem[Buchanan and Bryman(2018)]{Bryman2018}
D.A. Buchanan and A.~Bryman.
\newblock \emph{Unconventional Methodology in Organization and Management
  Research}, chapter Not another survey: The value of unconventional methods.
\newblock Oxford: Oxford University Press., 2018.
\newblock \doi{https://doi.org/10.1093/oso/9780198796978.003.0001}.

\bibitem[Casoli et~al.(2019)Casoli, Bonifazi, Ardizzone, Gravina, Russo,
  Sandulli, and Donnarumma]{Casoli2019}
Edoardo Casoli, Andrea Bonifazi, Giandomenico Ardizzone, Maria~Flavia Gravina,
  Giovanni~Fulvio Russo, Roberto Sandulli, and Luigia Donnarumma.
\newblock Comparative analysis of mollusc assemblages from different hard
  bottom habitats in the central tyrrhenian sea.
\newblock \emph{Diversity}, 11\penalty0 (5), 2019.
\newblock ISSN 1424-2818.
\newblock \doi{10.3390/d11050074}.
\newblock URL \url{https://www.mdpi.com/1424-2818/11/5/74}.

\bibitem[Chavez-Rosales et~al.(2019)Chavez-Rosales, Palka, Garrison, and
  Josephson]{chavez2019environmental}
Samuel Chavez-Rosales, Debra~L Palka, Lance~P Garrison, and Elizabeth~A
  Josephson.
\newblock Environmental predictors of habitat suitability and occurrence of
  cetaceans in the western north atlantic ocean.
\newblock \emph{Scientific Reports}, 9\penalty0 (1):\penalty0 1--11, 2019.

\bibitem[Clementi et~al.(2019)Clementi, Pistoia, Escudier, Delrosso, Drudi,
  Grandi, Lecci, Cret{\'\i}, Ciliberti, Coppini, Masina, and
  Pinardi]{Clementi2019}
E.~Clementi, J.~Pistoia, R.~Escudier, D.~Delrosso, M.~Drudi, A.~Grandi,
  R.~Lecci, S.~Cret{\'\i}, S.~Ciliberti, G.~Coppini, S.~Masina, and N.~Pinardi.
\newblock Mediterranean sea analysis and forecast (cmems med-currents, eas5
  system) [data set].
\newblock \emph{Copernicus Monitoring Environment Marine Service (CMEMS}, 2019.
\newblock
  \doi{https://doi.org/10.25423/CMCC/MEDSEA_ANALYSIS_FORECAST_PHY_006_013_EAS5}.

\bibitem[Corkeron et~al.(2011{\natexlab{a}})Corkeron, Minton, Collins, Findlay,
  Willson, and Baldwin]{corkeron2011spatial}
Peter~J Corkeron, Gianna Minton, Tim Collins, Ken Findlay, Andrew Willson, and
  Robert Baldwin.
\newblock Spatial models of sparse data to inform cetacean conservation
  planning: an example from oman.
\newblock \emph{Endangered Species Research}, 15\penalty0 (1):\penalty0 39--52,
  2011{\natexlab{a}}.

\bibitem[Corkeron et~al.(2011{\natexlab{b}})Corkeron, Minton, Collins, Findlay,
  Willson, and Baldwin]{Corkeron2011}
P.J. Corkeron, G.~Minton, T.~Collins, K.~Findlay, A.~Willson, and R.~Baldwin.
\newblock Spatial models of sparse data to inform cetacean conservation
  planning: an example from oman.
\newblock \emph{Endangered Species Research}, 15\penalty0 (1):\penalty0 39--52,
  2011{\natexlab{b}}.
\newblock URL \url{https://www.int-res.com/abstracts/esr/v15/n1/p39-52/}.

\bibitem[Dennis et~al.(2017)Dennis, Morgan, Freeman, Ridout, Brereton, Fox,
  Powney, and Roy]{Dennis_et_al2017}
E.~B. Dennis, B.J.~T. Morgan, S.~N. Freeman, M.S. Ridout, T.M. Brereton,
  R.~Fox, G.D. Powney, and D.~B. Roy.
\newblock Efficient occupancy model-fitting for extensive citizen-science data.
\newblock \emph{PLoS ONE}, 12\penalty0 (3), 2017.
\newblock \doi{https://doi.org/10.1371/journal.pone.0174433}.

\bibitem[Dorazio(2014)]{Dorazio2014}
Robert~M. Dorazio.
\newblock Accounting for imperfect detection and survey bias in statistical
  analysis of presence-only data.
\newblock \emph{Global Ecology and Biogeography}, 23\penalty0 (12):\penalty0
  1472--1484, 2014.
\newblock \doi{https://doi.org/10.1111/geb.12216}.
\newblock URL \url{https://onlinelibrary.wiley.com/doi/abs/10.1111/geb.12216}.

\bibitem[Fletcher~Jr. et~al.(2019)Fletcher~Jr., Hefley, Robertson, Zuckerberg,
  McCleery, and Dorazio]{Fletcher2019}
Robert~J. Fletcher~Jr., Trevor~J. Hefley, Ellen~P. Robertson, Benjamin
  Zuckerberg, Robert~A. McCleery, and Robert~M. Dorazio.
\newblock A practical guide for combining data to model species distributions.
\newblock \emph{Ecology}, 100\penalty0 (6):\penalty0 e02710, 2019.
\newblock \doi{https://doi.org/10.1002/ecy.2710}.
\newblock URL
  \url{https://esajournals.onlinelibrary.wiley.com/doi/abs/10.1002/ecy.2710}.

\bibitem[Florence et~al.(2020)Florence, Baudry, Pain, Sineau, and
  Pithon]{Florence2020.06.02.129536}
Matutini Florence, Jacques Baudry, Guillaume Pain, Morgane Sineau, and
  Jos{\'e}phine Pithon.
\newblock How citizen science could improve species distribution models and
  their independent assessment for conservation.
\newblock \emph{bioRxiv}, 2020.
\newblock \doi{10.1101/2020.06.02.129536}.
\newblock URL
  \url{https://www.biorxiv.org/content/early/2020/06/04/2020.06.02.129536}.

\bibitem[Fuglstad et~al.(2019)Fuglstad, Simpson, Lindgren, and
  Rue]{Fuglstad_etal2019}
Geir-Arne Fuglstad, Daniel Simpson, Finn Lindgren, and Håvard Rue.
\newblock Constructing priors that penalize the complexity of gaussian random
  fields.
\newblock \emph{Journal of the American Statistical Association}, 114\penalty0
  (525):\penalty0 445--452, 2019.
\newblock \doi{10.1080/01621459.2017.1415907}.
\newblock URL \url{https://doi.org/10.1080/01621459.2017.1415907}.

\bibitem[Hijmans(2019)]{raster2019}
Robert~J. Hijmans.
\newblock \emph{raster: Geographic Data Analysis and Modeling}, 2019.
\newblock URL \url{https://CRAN.R-project.org/package=raster}.
\newblock R package version 3.0-7.

\bibitem[{Horn}(1981)]{horn1981}
B.~K.~P. {Horn}.
\newblock Hill shading and the reflectance map.
\newblock \emph{Proceedings of the IEEE}, 69\penalty0 (1):\penalty0 14--47,
  1981.
\newblock \doi{10.1109/PROC.1981.11918}.

\bibitem[Illian(2019)]{HandbookEco2019}
Janine~B. Illian.
\newblock \emph{Handbook of Environmental and Ecological Statistics}, chapter
  Spatial and spatio-temporal point processes in ecological applications - 6,
  pages 97--132.
\newblock Handbooks of Modern Statistical Methods. Chapman and Hall - CRC Press
  Taylor \& Francis Group, 2019.

\bibitem[Isaac and Pocock(2015)]{Isaac2015}
Nick J.~B. Isaac and Michael J.~O. Pocock.
\newblock {Bias and information in biological records}.
\newblock \emph{Biological Journal of the Linnean Society}, 115\penalty0
  (3):\penalty0 522--531, 06 2015.
\newblock ISSN 0024-4066.
\newblock \doi{10.1111/bij.12532}.
\newblock URL \url{https://doi.org/10.1111/bij.12532}.

\bibitem[Isaac et~al.(2020)Isaac, Jarzyna, Keil, Dambly, Boersch-Supan,
  Browning, Freeman, Golding, Guillera-Arroita, Henrys, Jarvis, Lahoz-Monfort,
  Pagel, Pescott, Schmucki, Simmonds, and O'Hara]{Isaac2020}
Nick~J.B. Isaac, Marta~A. Jarzyna, Petr Keil, Lea~I. Dambly, Philipp~H.
  Boersch-Supan, Ella Browning, Stephen~N. Freeman, Nick Golding, Gurutzeta
  Guillera-Arroita, Peter~A. Henrys, Susan Jarvis, Jos{\'e} Lahoz-Monfort,
  J{\"o}rn Pagel, Oliver~L. Pescott, Reto Schmucki, Emily~G. Simmonds, and
  Robert~B. O'Hara.
\newblock Data integration for large-scale models of species distributions.
\newblock \emph{Trends in Ecology and Evolution}, 35\penalty0 (1):\penalty0 56
  -- 67, 2020.
\newblock ISSN 0169-5347.
\newblock \doi{https://doi.org/10.1016/j.tree.2019.08.006}.
\newblock URL
  \url{http://www.sciencedirect.com/science/article/pii/S0169534719302551}.

\bibitem[ISPRA()]{FLTNET}
ISPRA.
\newblock Fixed line transect mediterranean monitoring network.
\newblock URL
  \url{https://www.isprambiente.gov.it/en/activities/biodiversity/flt-mediterranean-monitoring-network-marine-species-and-threats}.

\bibitem[ISPRA(2016.)]{ISPRA2016}
ISPRA.
\newblock Fixed line transect monitoring using ferries as platform of
  observation for marine mega and macro fauna and main threats. monitoring
  protocol for cetaceans and sea turtles.
\newblock ISPRA Agreement - Technical annex~1, ISPRA., pp.19, 2016.

\bibitem[Kery et~al.(2010)Kery, Royle, Schmid, Schaub, Volet, Haflinger, and
  Zbinden]{Kery_et_al2010}
Marc Kery, J.~A. Royle, H.~Schmid, M.~Schaub, B.~Volet, G.~Haflinger, and
  N.~Zbinden.
\newblock Site-occupancy distribution modeling to correct population-trend
  estimates derived from opportunistic observations.
\newblock \emph{Conservation Biology}, 24\penalty0 (5):\penalty0 1388--1397,
  2010.
\newblock \doi{https://doi.org/10.1111/j.1523-1739.2010.01479.x}.
\newblock URL
  \url{https://conbio.onlinelibrary.wiley.com/doi/abs/10.1111/j.1523-1739.2010.01479.x}.

\bibitem[Lindgren et~al.(2011)Lindgren, Rue, and Lindström]{lindgrenSPDE}
Finn Lindgren, Håvard Rue, and Johan Lindström.
\newblock An explicit link between gaussian fields and gaussian markov random
  fields: the stochastic partial differential equation approach.
\newblock \emph{Journal of the Royal Statistical Society: Series B (Statistical
  Methodology)}, 73\penalty0 (4):\penalty0 423--498, 2011.
\newblock \doi{https://doi.org/10.1111/j.1467-9868.2011.00777.x}.
\newblock URL
  \url{https://rss.onlinelibrary.wiley.com/doi/abs/10.1111/j.1467-9868.2011.00777.x}.

\bibitem[Mart{\'\i}n~M{\'\i}guez et~al.(2019)Mart{\'\i}n~M{\'\i}guez,
  Novellino, Vinci, Claus, Calewaert, Vallius, Schmitt, Pititto, Giorgetti,
  Askew, Iona, Schaap, Pinardi, Harpham, Kater, Populus, She, Palazov, McMeel,
  Oset, Lear, Manzella, Gorringe, Simoncelli, Larkin, Holdsworth, Arvanitidis,
  Molina~Jack, Chaves~Montero, Herman, and Hernandez]{EMODNET2019}
Bel{\'e}n Mart{\'\i}n~M{\'\i}guez, Antonio Novellino, Matteo Vinci, Simon
  Claus, Jan-Bart Calewaert, Henry Vallius, Thierry Schmitt, Alessandro
  Pititto, Alessandra Giorgetti, Natalie Askew, Sissy Iona, Dick Schaap, Nadia
  Pinardi, Quillon Harpham, Belinda~J. Kater, Jacques Populus, Jun She,
  Atanas~Vasilev Palazov, Oonagh McMeel, Paula Oset, Dan Lear, Giuseppe M.~R.
  Manzella, Patrick Gorringe, Simona Simoncelli, Kate Larkin, Neil Holdsworth,
  Christos~Dimitrios Arvanitidis, Maria~Eugenia Molina~Jack, Maria del~Mar
  Chaves~Montero, Peter M.~J. Herman, and Francisco Hernandez.
\newblock The european marine observation and data network (emodnet): Visions
  and roles of the gateway to marine data in europe.
\newblock \emph{Frontiers in Marine Science}, 6:\penalty0 313, 2019.
\newblock ISSN 2296-7745.
\newblock \doi{10.3389/fmars.2019.00313}.
\newblock URL
  \url{https://www.frontiersin.org/article/10.3389/fmars.2019.00313}.

\bibitem[Meissner et~al.(2012)Meissner, MacLeod, Richard, Ridoux, and
  Pierce]{meissner2012}
A.M. Meissner, C.D. MacLeod, P.~Richard, V.~Ridoux, and G.~Pierce.
\newblock Feeding ecology of striped dolphins, stenella coeruleoalba, in the
  north-western mediterranean sea based on stable isotope analyses.
\newblock \emph{Journal of the Marine Biological Association of the United
  Kingdom}, 92\penalty0 (8):\penalty0 1677--1687, 2012.
\newblock \doi{10.1017/S0025315411001457}.

\bibitem[Mikula and Tryjanowski(2016)]{mikula2016internet}
Peter Mikula and Piotr Tryjanowski.
\newblock Internet searching of bird-bird associations: A case of bee-eaters
  hitchhiking large african birds.
\newblock \emph{Biodiversity Observations}, pages 1--6, 2016.

\bibitem[Miller et~al.(2019)Miller, Pacifici, Sanderlin, and Reich]{Miller2019}
David A.~W. Miller, Krishna Pacifici, Jamie~S. Sanderlin, and Brian~J. Reich.
\newblock The recent past and promising future for data integration methods to
  estimate species' distributions.
\newblock \emph{Methods in Ecology and Evolution}, 10\penalty0 (1):\penalty0
  22--37, 2019.
\newblock \doi{https://doi.org/10.1111/2041-210X.13110}.
\newblock URL
  \url{https://besjournals.onlinelibrary.wiley.com/doi/abs/10.1111/2041-210X.13110}.

\bibitem[Monsarrat et~al.(2019)Monsarrat, Boshoff, and Kerley]{monserrat2019}
Sophie Monsarrat, Andre~F. Boshoff, and Graham I.~H. Kerley.
\newblock Accessibility maps as a tool to predict sampling bias in historical
  biodiversity occurrence records.
\newblock \emph{Ecography}, 42\penalty0 (1):\penalty0 125--136, 2019.
\newblock \doi{https://doi.org/10.1111/ecog.03944}.
\newblock URL \url{https://onlinelibrary.wiley.com/doi/abs/10.1111/ecog.03944}.

\bibitem[Pace et~al.(2014)Pace, Mussi, Gordon, and W{\"u}rtz]{Paceetal2014}
Daniela~S. Pace, Barbara Mussi, Jonathan C.~D. Gordon, and Maurizio W{\"u}rtz.
\newblock Foreword.
\newblock \emph{Aquatic Conservation: Marine and Freshwater Ecosystems},
  24\penalty0 (S1):\penalty0 1--3, 2014.
\newblock \doi{https://doi.org/10.1002/aqc.2457}.
\newblock URL \url{https://onlinelibrary.wiley.com/doi/abs/10.1002/aqc.2457}.

\bibitem[Pace et~al.(2012)Pace, Pulcini, and Triossi]{Paceetal2012}
Daniela~Silvia Pace, Marina Pulcini, and Francesca Triossi.
\newblock {Anthropogenic food patches and association patterns of Tursiops
  truncatus at Lampedusa island, Italy}.
\newblock \emph{Behavioral Ecology}, 23\penalty0 (2):\penalty0 254--264, 2012.
\newblock ISSN 1045-2249.
\newblock \doi{10.1093/beheco/arr180}.
\newblock URL \url{https://doi.org/10.1093/beheco/arr180}.

\bibitem[Pace et~al.(2018)Pace, Arcangeli, Mussi, Vivaldi, Ledon, Lagorio,
  Giacomini, Pavan, and Ardizzone]{pace2018}
Daniela~Silvia Pace, Antonella Arcangeli, Barbara Mussi, Carlotta Vivaldi,
  Cristina Ledon, Serena Lagorio, Giancarlo Giacomini, Gianni Pavan, and
  Giandomenico Ardizzone.
\newblock Habitat suitability modeling in different sperm whale social groups.
\newblock \emph{The Journal of Wildlife Management}, 82\penalty0 (5):\penalty0
  1062--1073, 2018.

\bibitem[Pace et~al.(2019)Pace, Giacomini, Campana, Paraboschi, Pellegrino,
  Silvestri, Alessi, Angeletti, Cafaro, Pavan, Ardizzone, and
  Arcangeli]{Paceetal2019}
Daniela~Silvia Pace, Giancarlo Giacomini, Ilaria Campana, Miriam Paraboschi,
  Giuliana Pellegrino, Margherita Silvestri, Jessica Alessi, Dario Angeletti,
  Valentina Cafaro, Gianni Pavan, Giandomenico Ardizzone, and Antonella
  Arcangeli.
\newblock An integrated approach for cetacean knowledge and conservation in the
  central mediterranean sea using research and social media data sources.
\newblock \emph{Aquatic Conservation: Marine and Freshwater Ecosystems},
  29\penalty0 (8):\penalty0 1302--1323, 2019.
\newblock \doi{https://doi.org/10.1002/aqc.3117}.
\newblock URL \url{https://onlinelibrary.wiley.com/doi/abs/10.1002/aqc.3117}.

\bibitem[Pace et~al.(2021)Pace, Mussi, Vella, and Vella]{Paceetal2021}
D.S. Pace, B.~Mussi, J.~Vella, and A.~Vella.
\newblock Facts and outcomes of the mediterranean short-beaked common dolphin
  ($delphinus$ $delphis$) workshop.
\newblock \emph{Aquatic Conservation: Marine and Freshwater Ecosystems},
  n/a\penalty0 (n/a), 2021.
\newblock \doi{https://doi.org/10.1002/aqc.3549}.
\newblock URL \url{https://onlinelibrary.wiley.com/doi/abs/10.1002/aqc.3549}.

\bibitem[Phillips et~al.(2006)Phillips, Anderson, and Schapire]{maxent2006}
Steven~J. Phillips, Robert~P. Anderson, and Robert~E. Schapire.
\newblock Maximum entropy modeling of species geographic distributions.
\newblock \emph{Ecological Modelling}, 190\penalty0 (3):\penalty0 231--259,
  2006.
\newblock ISSN 0304-3800.
\newblock \doi{https://doi.org/10.1016/j.ecolmodel.2005.03.026}.
\newblock URL
  \url{https://www.sciencedirect.com/science/article/pii/S030438000500267X}.

\bibitem[Pulcini et~al.(2014)Pulcini, Pace, Manna, Triossi, and
  Fortuna]{Pulcinietal2014}
M.~Pulcini, D.~Pace, G.~Manna, Francesca Triossi, and C.~Fortuna.
\newblock Distribution and abundance estimates of bottlenose dolphins (
  tursiops truncatus ) around lampedusa island (sicily channel, italy):
  implications for their management.
\newblock \emph{Journal of the Marine Biological Association of the United
  Kingdom}, 94:\penalty0 1175--1184, 2014.
\newblock \doi{doi:10.1017/S0025315413000842}.

\bibitem[Redfern et~al.(2006)Redfern, Ferguson, Becker, Hyrenbach, Good,
  Barlow, Kaschner, Baumgartner, Forney, Ballance, Fauchald, Halpin, Hamazaki,
  Pershing, Qian, Read, Reilly, Torres, and Werner]{Redfern2006}
J.~V. Redfern, M.~C. Ferguson, E.~A. Becker, K.~D. Hyrenbach, C.~Good,
  J.~Barlow, K.~Kaschner, M.~F. Baumgartner, K.~A. Forney, L.~T. Ballance,
  P.~Fauchald, P.~Halpin, T.~Hamazaki, A.~J. Pershing, S.~S. Qian, A.~Read,
  S.~B. Reilly, L.~Torres, and F.~Werner.
\newblock Techniques for cetacean\&{\#}150;habitat modeling.
\newblock \emph{Marine Ecology Progress Series}, 310:\penalty0 271--295, 2006.
\newblock URL \url{https://www.int-res.com/abstracts/meps/v310/p271-295/}.

\bibitem[Redfern et~al.(2017)Redfern, Moore, Fiedler, de~Vos, Brownell~Jr,
  Forney, Becker, and Ballance]{Redfern2017}
Jessica~V. Redfern, Thomas~J. Moore, Paul~C. Fiedler, Asha de~Vos, Robert~L.
  Brownell~Jr, Karin~A. Forney, Elizabeth~A. Becker, and Lisa~T. Ballance.
\newblock Predicting cetacean distributions in data-poor marine ecosystems.
\newblock \emph{Diversity and Distributions}, 23\penalty0 (4):\penalty0
  394--408, 2017.
\newblock \doi{https://doi.org/10.1111/ddi.12537}.
\newblock URL \url{https://onlinelibrary.wiley.com/doi/abs/10.1111/ddi.12537}.

\bibitem[Rue and Held(2005)]{RueHeld:2005}
Havard Rue and Leonhard Held.
\newblock \emph{Gaussian Markov Random Fields. Theory and Applications}.
\newblock Chapman \& Hall/CRC, 2005.

\bibitem[Rue et~al.(2009)Rue, Martino, and Chopin]{rueetal2009}
Håvard Rue, Sara Martino, and Nicolas Chopin.
\newblock Approximate bayesian inference for latent gaussian models by using
  integrated nested laplace approximations.
\newblock \emph{Journal of the Royal Statistical Society: Series B (Statistical
  Methodology)}, 71\penalty0 (2):\penalty0 319--392, 2009.
\newblock \doi{https://doi.org/10.1111/j.1467-9868.2008.00700.x}.
\newblock URL
  \url{https://rss.onlinelibrary.wiley.com/doi/abs/10.1111/j.1467-9868.2008.00700.x}.

\bibitem[Salon et~al.(2019)Salon, Cossarini, Bolzon, Feudale, Lazzari, Teruzzi,
  Solidoro, and Crise]{Salon2019}
Stefano Salon, Gianpiero Cossarini, Giorgio Bolzon, Laura Feudale, Paolo
  Lazzari, Anna Teruzzi, Cosimo Solidoro, and Alessandro Crise.
\newblock Novel metrics based on biogeochemical argo data to improve the model
  uncertainty evaluation of the cmems mediterranean marine ecosystem forecasts.
\newblock \emph{Ocean Science}, 15\penalty0 (4):\penalty0 997--1022, 2019.

\bibitem[Santamaria et~al.(2017)Santamaria, Alvarez, Greidanus, Syrris, Soille,
  and Argentieri]{remotesensing2017}
Carlos Santamaria, Marlene Alvarez, Harm Greidanus, Vasileios Syrris, Pierre
  Soille, and Pietro Argentieri.
\newblock Mass processing of sentinel-1 images for maritime surveillance.
\newblock \emph{Remote Sensing}, 9\penalty0 (7), 2017.
\newblock ISSN 2072-4292.
\newblock \doi{10.3390/rs9070678}.
\newblock URL \url{https://www.mdpi.com/2072-4292/9/7/678}.

\bibitem[Sicacha-Parada et~al.(2020)Sicacha-Parada, Steinsland, Cretois, and
  Borgelt]{Sicachaparada2020}
Jorge Sicacha-Parada, Ingelin Steinsland, Benjamin Cretois, and Jan Borgelt.
\newblock Accounting for spatial varying sampling effort due to accessibility
  in citizen science data: A case study of moose in norway.
\newblock \emph{Spatial Statistics}, page 100446, 2020.
\newblock ISSN 2211-6753.
\newblock \doi{https://doi.org/10.1016/j.spasta.2020.100446}.
\newblock URL
  \url{http://www.sciencedirect.com/science/article/pii/S2211675320300403}.

\bibitem[Simoncelli et~al.(2019)Simoncelli, Fratianni, Pinardi, Grandi, Drudi,
  Oddo, and Dobricic]{Simoncelli2019}
S.~Simoncelli, C.~Fratianni, N.~Pinardi, A.~Grandi, M.~Drudi, P.~Oddo, and
  S.~Dobricic.
\newblock Mediterranean sea physical reanalysis (cmems med-physics)[data set].
\newblock \emph{Copernicus Monitoring Environment Marine Service (CMEMS)},
  2019.
\newblock \doi{https://doi.org/10.25423/MEDSEA_REANALYSIS_PHYS_006_004}.

\bibitem[Simpson et~al.(2016)Simpson, Illian, Lindgren, Sørbye, and
  Rue]{simpsonetal2016}
D.~Simpson, J.~B. Illian, F.~Lindgren, S.~H. Sørbye, and H.~Rue.
\newblock {Going off grid: computationally efficient inference for log-Gaussian
  Cox processes}.
\newblock \emph{Biometrika}, 103\penalty0 (1):\penalty0 49--70, 02 2016.
\newblock ISSN 0006-3444.
\newblock \doi{10.1093/biomet/asv064}.
\newblock URL \url{https://doi.org/10.1093/biomet/asv064}.

\bibitem[Soranno and Schimel(2014)]{soranno2014macrosystems}
Patricia~A Soranno and David~S Schimel.
\newblock Macrosystems ecology: big data, big ecology, 2014.

\bibitem[Stephenson et~al.(2020)Stephenson, Goetz, Sharp, Mouton, Beets,
  Roberts, MacDiarmid, Constantine, and Lundquist]{Stephenson2020}
Fabrice Stephenson, Kimberly Goetz, Ben~R. Sharp, Théophile~L. Mouton,
  Fenna~L. Beets, Jim Roberts, Alison~B. MacDiarmid, Rochelle Constantine, and
  Carolyn~J. Lundquist.
\newblock Modelling the spatial distribution of cetaceans in new zealand
  waters.
\newblock \emph{Diversity and Distributions}, 26\penalty0 (4):\penalty0
  495--516, 2020.
\newblock \doi{https://doi.org/10.1111/ddi.13035}.
\newblock URL \url{https://onlinelibrary.wiley.com/doi/abs/10.1111/ddi.13035}.

\bibitem[Teruzzi et~al.(2019)Teruzzi, Bolzon, Cossarini, Lazzari, Salon, Crise,
  and Solidoro]{Teruzzi2019}
A.~Teruzzi, G.~Bolzon, G.~Cossarini, P.~Lazzari, S.~Salon, A.~Crise, and
  C.~Solidoro.
\newblock Mediterranean sea biogeochemical reanalysis (cmems
  med-biogeochemistry)[data set].
\newblock \emph{Copernicus Monitoring Environment Marine Service (CMEMS)},
  2019.
\newblock \doi{https://doi.org/10.25423/MEDSEA_REANALYSIS_BIO_006_008}.

\bibitem[Thompson and Ramsey(1987)]{Thompson1987}
Steven~K. Thompson and Fred~L. Ramsey.
\newblock Detectability functions in observing spatial point processes.
\newblock \emph{Biometrics}, 43\penalty0 (2):\penalty0 355--362, 1987.
\newblock ISSN 0006341X, 15410420.
\newblock URL \url{http://www.jstor.org/stable/2531818}.

\bibitem[Triossi et~al.(2013)Triossi, Willis, and Pace]{Triossietal2013}
Francesca Triossi, Trevor~J. Willis, and Daniela~S. Pace.
\newblock Occurrence of bottlenose dolphins tursiops truncatus in natural gas
  fields of the northwestern adriatic sea.
\newblock \emph{Marine Ecology}, 34\penalty0 (3):\penalty0 373--379, 2013.
\newblock \doi{https://doi.org/10.1111/maec.12020}.
\newblock URL \url{https://onlinelibrary.wiley.com/doi/abs/10.1111/maec.12020}.

\bibitem[Vella et~al.(2021)Vella, Murphy, Giménez, de~Stephanis, Mussi, Vella,
  Larbi~Doukara, and Pace]{Vella2021common}
Adriana Vella, Sinéad Murphy, Joan Giménez, Renaud de~Stephanis, Barbara
  Mussi, Joseph~G. Vella, Kamel Larbi~Doukara, and Daniela~Silvia Pace.
\newblock The conservation of the endangered mediterranean common dolphin
  ($delphinus$ $delphis$): Current knowledge and research priorities.
\newblock \emph{Aquatic Conservation: Marine and Freshwater Ecosystems}, pages
  1--27, 2021.

\bibitem[Ventura et~al.(2015)Ventura, Jona~Lasinio, and Ardizzone]{Ventura2015}
Daniele Ventura, Giovanna Jona~Lasinio, and Giandomenico Ardizzone.
\newblock Temporal partitioning of microhabitat use among four juvenile fish
  species of the genus diplodus (pisces: Perciformes, sparidae).
\newblock \emph{Marine Ecology}, 36\penalty0 (4):\penalty0 1013--1032, 2015.
\newblock \doi{https://doi.org/10.1111/maec.12198}.
\newblock URL \url{https://onlinelibrary.wiley.com/doi/abs/10.1111/maec.12198}.

\bibitem[Yuan et~al.(2017)Yuan, Bachl, Lindgren, Borchers, Illian, Buckland,
  Rue, and Gerrodette]{yuan2017}
Yuan Yuan, Fabian~E. Bachl, Finn Lindgren, David~L. Borchers, Janine~B. Illian,
  Stephen~T. Buckland, Håvard Rue, and Tim Gerrodette.
\newblock Point process models for spatio-temporal distance sampling data from
  a large-scale survey of blue whales.
\newblock \emph{Ann. Appl. Stat.}, 11\penalty0 (4):\penalty0 2270--2297, 12
  2017.
\newblock \doi{10.1214/17-AOAS1078}.
\newblock URL \url{https://doi.org/10.1214/17-AOAS1078}.

\end{thebibliography}

\newpage


\section*{Supplementary Material}

\appendix
\section{Environmental Data Sources and Management}
\begin{figure}%
    \centering
    \subfloat[]{
        \includegraphics[width=0.45\textwidth]{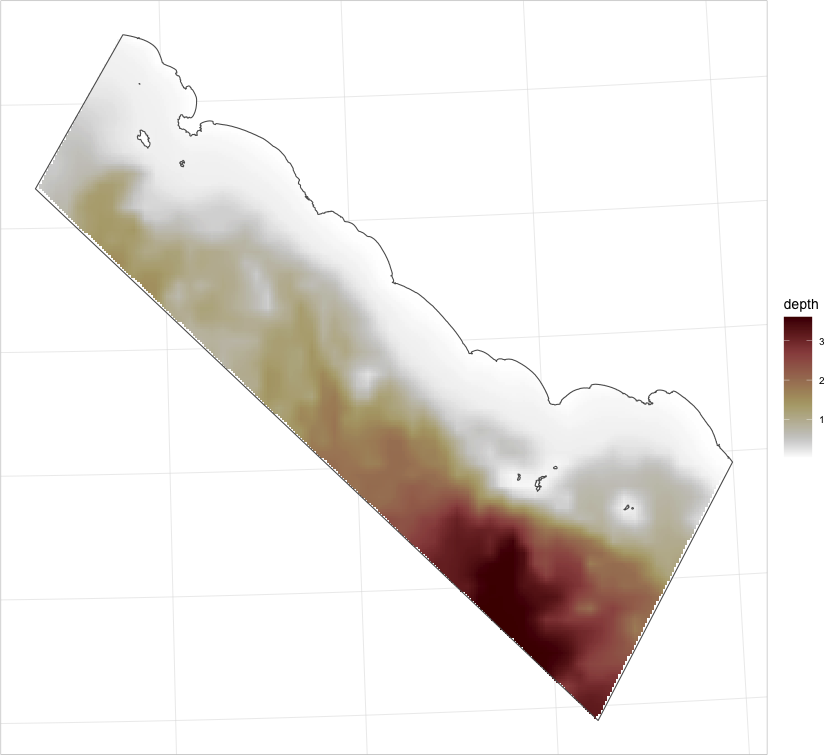}
    }%
    \qquad
    \subfloat[]{
        \includegraphics[width=0.45\textwidth]{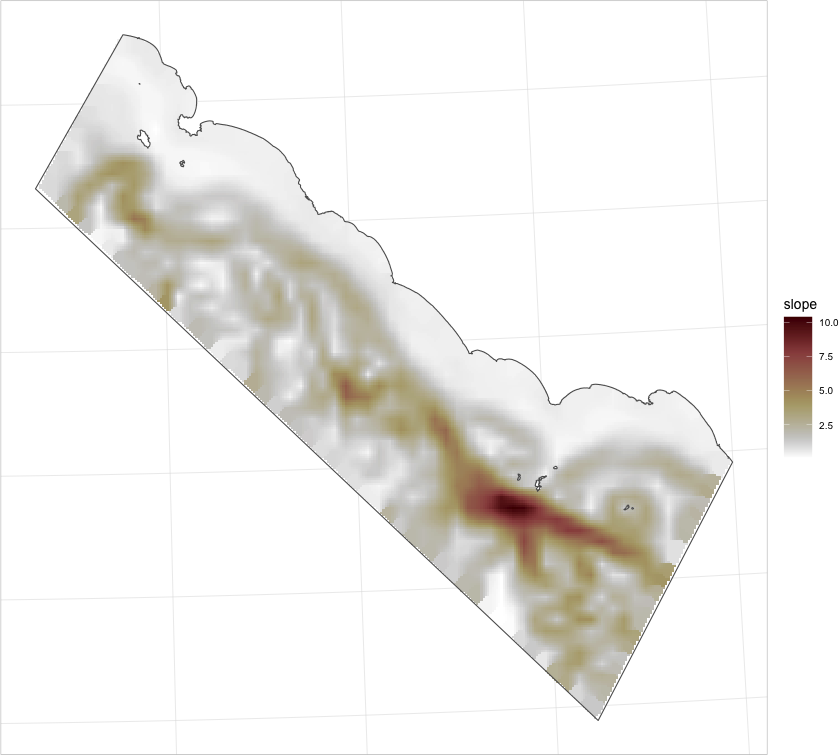}
    }%
    \qquad
    \caption{Map of depth (a) and slope (b) in the area of interest.}\label{fig:depth_slope}%
\end{figure}

We used temperature, primary productivity, slope, and depth as potential covariates.  
\begin{itemize}
\item \textbf{Temperature and primary productivity} were retrieved from COPERNICUS platform. COPERNICUS is part of the European Union’s Earth Observation Programme and constitutes an institutional platform collecting and providing environmental data for the ultimate benefit of all European citizens. It is based on both InSitu and satellite monitoring. The Programme is coordinated and managed by the European Commission and benefits of essential partnerships with the Member States, the European Space Agency (ESA), the European Organization for the Exploitation of Meteorological Satellites (EUMETSAT), the European Centre for Medium-Range Weather Forecasts (ECMWF), EU Agencies and Mercator Océan. For our purposes, we downloaded the oceanographic data from the CMEMS (Copernicus Marine Environment Monitoring Service - \url{https://marine.copernicus.eu/}. 
\item \textbf{Depth} data (Figure \ref{fig:depth_slope}a) were downloaded from GEBCO (General bathymetric Chart of the Ocean - \url{https://www.gebco.net/}) which provides the most authoritative publicly-available bathymetry of the world’s oceans. GEBCO\_2020 Grid has a spatial resolution of 15 arc seconds, corresponding to about 350 m at 41° latitude. 
\item \textbf{Slope} (Figure \ref{fig:depth_slope}b) was computed from a smoothed Depth surface (resolution 1km$\times$1 km) data using \cite{horn1981} algorithm (see the function \texttt{terrain} in library \texttt{raster} by \citet{raster2019}). Obtained values ranges are in $[0,10)$ degrees.
\end{itemize}
Envaronmental variables details are given below:
\begin{itemize}
\item Temperature, years 2008 - 2018, with a grid resolution of 0.063 decimal degrees,   72 depth levels,  the name of the service according to Copernicus is MEDSEA\_REANALYSIS\_PHYS\_006\_004, see \citet{Simoncelli2019} 
\item Temperature, year 2019, with a grid resolution of 0.042 decimal degrees, 141 depth levels, the name of the service according to Copernicus is  MEDSEA\_ANALYSIS\_FORECAST\_PHY\_006\_013, see  \citet{Clementi2019} 
\item Primary Productivity, years 2008 - 2018, with a grid resolution of 0.063 decimal degrees,  72 depth levels,  the name of the service according to Copernicus is  MEDSEA\_REANALYSIS\_BIO\_006\_008, see  \citet{Teruzzi2019}.
\item Primary Productivity, year 2019,  with a grid resolution of 0.042,  141 depth levels,  the name of the service according to Copernicus is  MEDSEA\_ANALYSIS\_FORECAST\_BIO\_006\_014, see \citet{Bolzon2020,Salon2019}.
\end{itemize}

Environmental data were then aligned to the estimation grid both in terms of spatial and temporal resolution. First, data were re-projected using UTM coordinates (EPSG:32632). Depth raster was up-scaled to the 1km x 1km estimation grid resolution using a Median Smoothing. Temperature and primary productivity data required more intense management than depth. In the details given above, both primary productivity and temperature datasets were characterized by differences in spatial resolution and the number of depth levels between the temporal bin “2008-2018” and data collected in 2019. Hence, during the download data, filtered, keeping just the more superficial layer. The latter ranges between 0 and -1.47 m for the “2008-2018” series and between 0 and -1.01 m for the “2019” series. Primary productivity data were first harmonized with the same measurement unit (from moles/m3/s to mg/m3/day). Variables were then down-scaled using Generalized Additive Models. We fitted a GAM to each stack raster layer, that is, each month of each year of the investigated temporal range, and then predicted the variable value (i.e. temperature or primary productivity) over a surface gridded at 1 km x 1km. Predictions were, aggregated by average to obtain one value for each cell. 
All variables are available from Copernicus as monthly means.

\section{SM data detection function: Estimation of the intensity from observations of all species}
\label{supp_section:logintensity}

We assume that the location of observations of all species in the SM dataset (Figure \ref{fig:area_interest}(b) )  are properly described by a LGCP point process with log-intensity
\begin{equation}
    \log \gamma(s) = \alpha + u(s)
\end{equation}
where $\alpha$ is a common intercept and $u(s)$ is a zero-mean Gaussian process with Matern covariance function of order 1 with range $\rho_u$ and standard deviation $\sigma_u$.
Here also, we follow the SPDE approach as in \cite{yuan2017} to represent the Gaussian field and use the same mesh and the same priors for $\rho_u$  and $\sigma_u$ as in the main model.
\begin{figure}%
    \centering
  
        \includegraphics[width=0.4\textwidth]{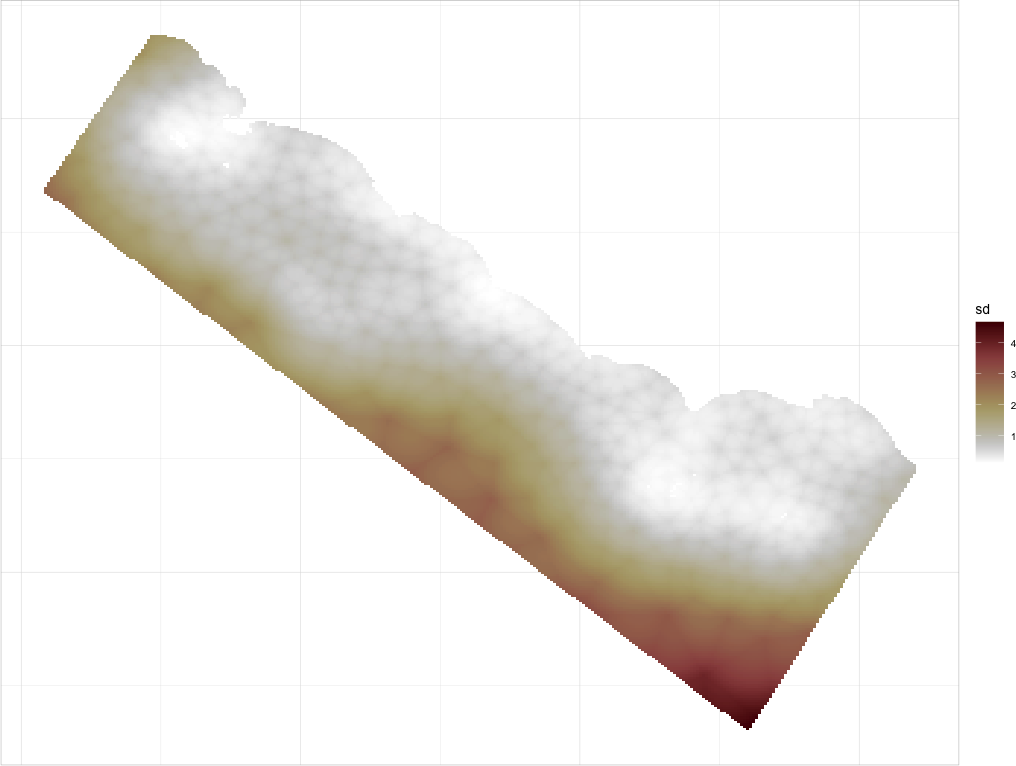}
  
    \caption{Estimated standard deviation of $\log(\gamma(s))$.}%
    \label{supp_fig:sd_intensity}

\end{figure}
The estimated log-intensity surface is shown in Figure~\ref{fig:int_surfaces}c while its standard deviation is shown in Figure \ref{supp_fig:sd_intensity}). Table \ref{tab_supp:intensity} reports the posterior mean and 95\% credible interval for $\rho_u$  and $\sigma_u$. 
\begin{table}[ht]
\centering
\begin{tabular}{rrrr}
  \hline
 & mean & 2.5\% & 97.5\% \\ 
  \hline
$\rho_u$  & 114.23 & 73.68 & 172.84 \\ 
$\sigma_u$   & 3.34 & 2.27 & 4.82 \\ 
   \hline
\end{tabular}
\caption{Posterior mean and 05\% credible interval for $\rho_u$  and $\sigma_u$} 
\label{tab_supp:intensity}
\end{table}

\section{ Shape of the detection functions}

One of the main ideas in this work is to correct the social media dataset for biases due to the larger number of sightings expected in areas that are more visited by small boats (the primary source of the SM data). We use three different proxies for the small boat density: the distance from the coastline $d_1(s)$, the log-density of vessels from the EMODnet data $d_2(s)$, and the estimated log intensity for all observations in the SM dataset $d_3(s)$. 

We have computed the the three measures for the
862 location where a dolphin was observed (283 striped and 579 bottlenose dolphins) and for about 32000 points evenly distributed on a grid over the whole area of interest. 
\begin{figure}%
    \centering
    \subfloat[]{
        \includegraphics[width=0.33\textwidth]{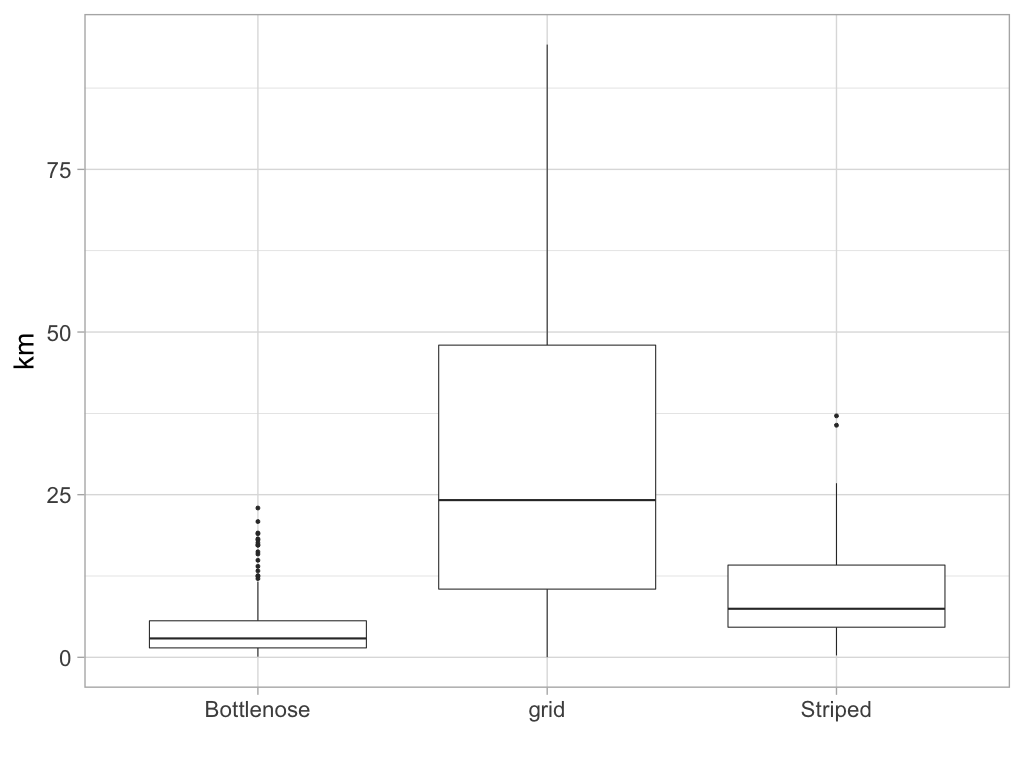}
    }%
    \subfloat[]{
        \includegraphics[width=0.33\textwidth]{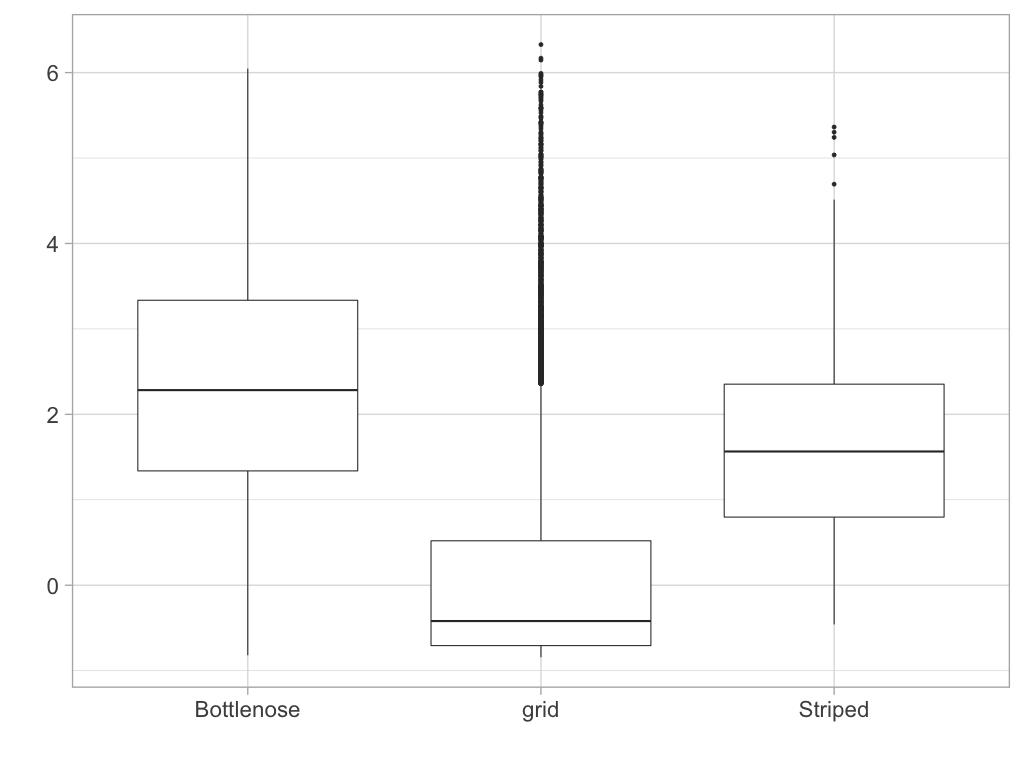}
    }%
    \subfloat[]{
        \includegraphics[width=0.33\textwidth]{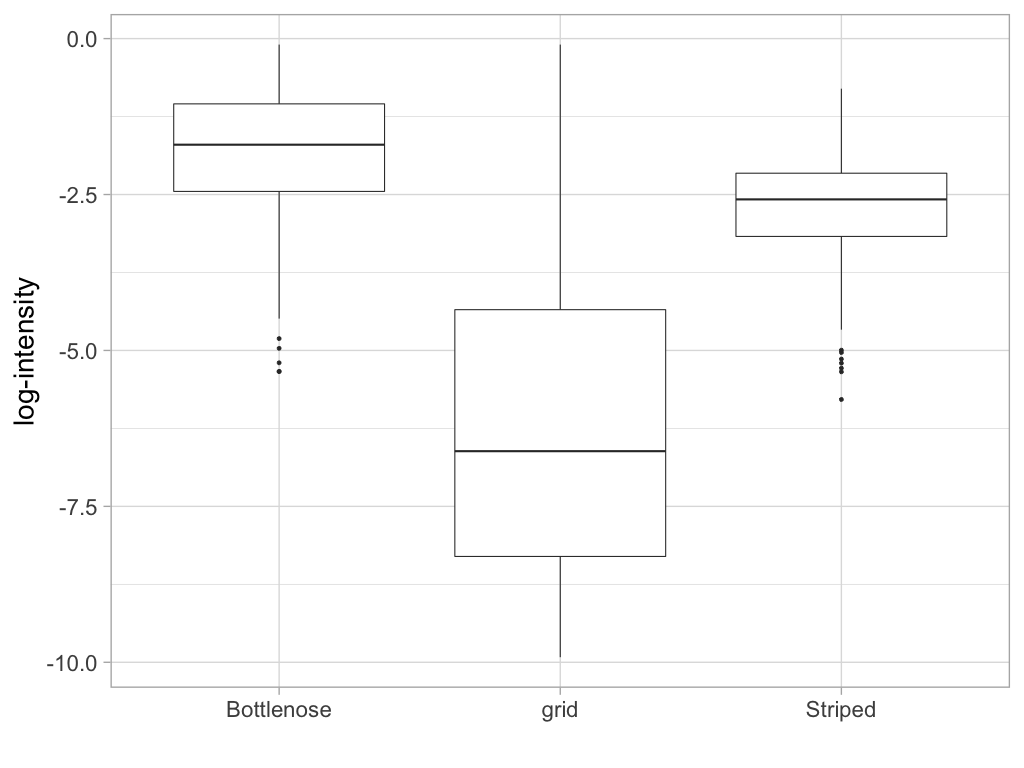}
    }%
    \caption{(a) Boxplot of distance from the coastline, (b) log vessel density and (c) estimated log-intensity from all observations in the SM dataset. Center: Dense grid of about 21 thousand points. Left: 579 locations with observed bottlenose Right: 283 locations with observed striped dolphins. }%
    \label{supp_fig:boxplot}%
\end{figure}
The boxplots of these measures for each set of points are displayed in Fig. \ref{supp_fig:boxplot}. In all three cases, both species' measures appear to come from a different distribution than the measures for the grid. A Kolmogorov–Smirnov test was performed to determine if the grid data and the bottlenose and striped dolphin data follow the same distribution or not. In all cases, the result ($p-value<2e-16$) indicates that the sets of measures do not follow the same distribution. That is an indication of a non-random sampling process. 

We then explore the relationship between each of the proxies $d_i(s)$, $i = 1,2,3$ and, the realative probability $q(s_{d_i})$ of retaining a point in a location with measure  $d_i(s)$ (i.e. not thinning) in the observed pattern. To estimate this, we grouped all three measures for striped and bottlenose sightings locations and for the grid points into 100 bins. For each bin we computed 
\begin{equation}
    \hat{q}^j(s_{d_i}) = \frac{\hat{p}^j_{obs}(s_{d_i})}{\hat{p}_{grid}(s_{d_i})}
\end{equation}
with $\hat{p}^j_{obs}$, $j = 1,2$ and $\hat{p}_{grid}$, the proportion of points  that are part of the bin in the observed pattern of the two species and the dense grid, respectively. 
\begin{figure}%
    \centering
    \subfloat[]{
        \includegraphics[width=0.33\textwidth]{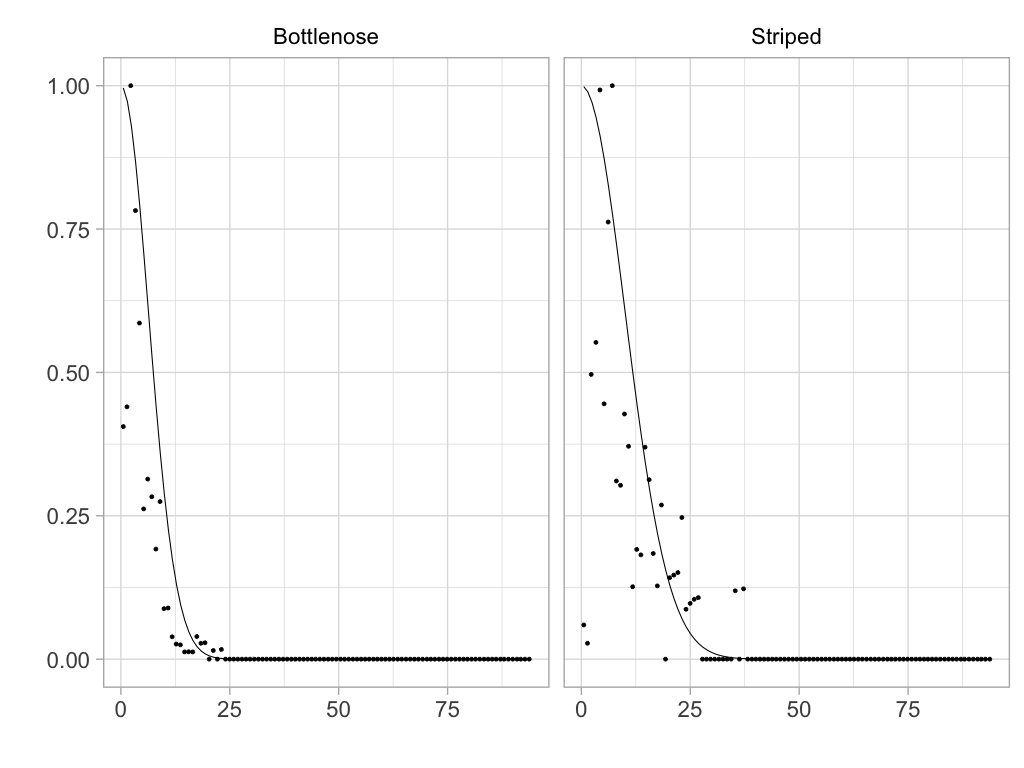}
    }%
    \subfloat[]{
        \includegraphics[width=0.33\textwidth]{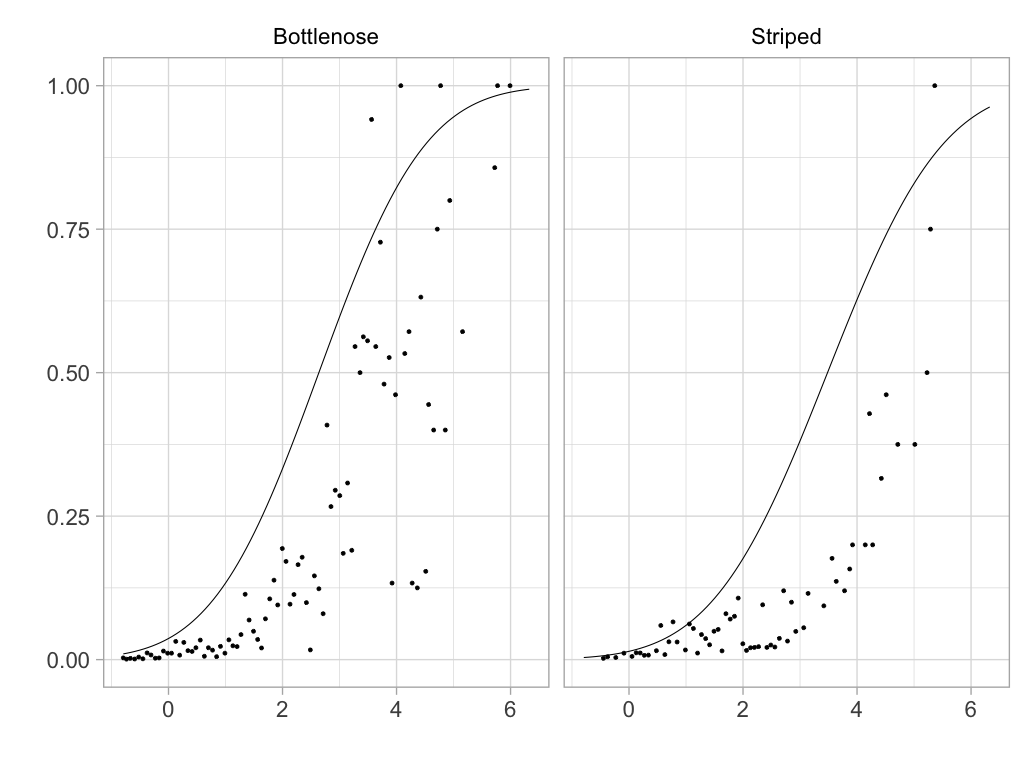}
    }%
    \subfloat[]{
        \includegraphics[width=0.33\textwidth]{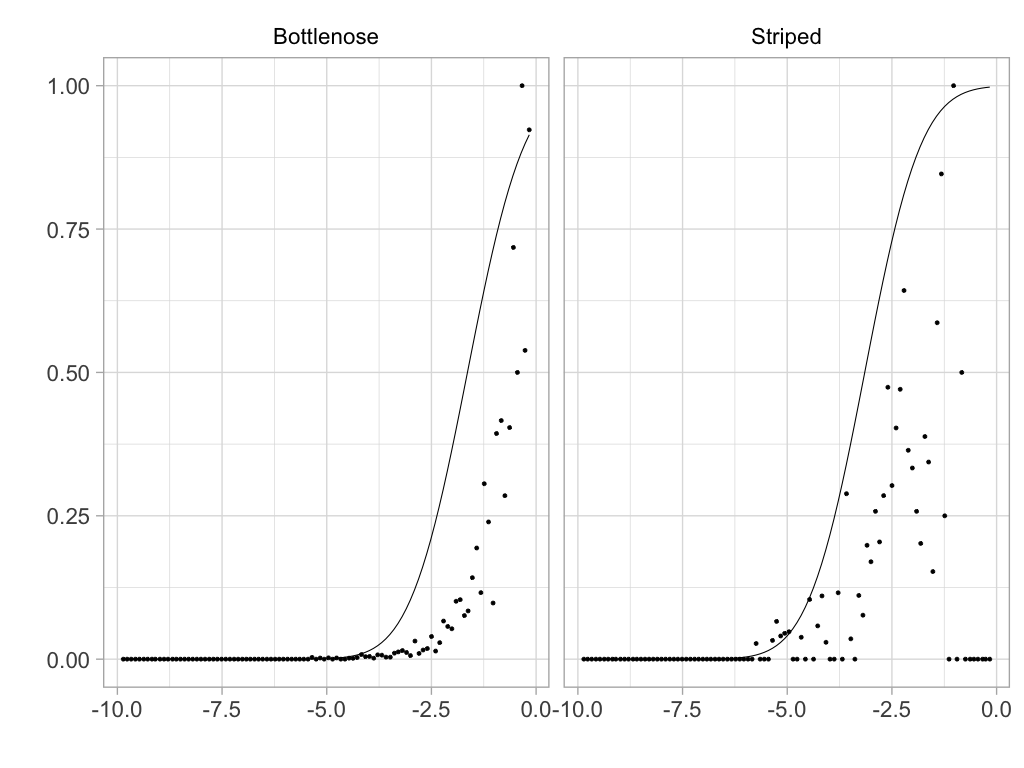}
    }%
    \caption{Relationship between the observed ratio $\hat{q}^j(s_{d_i})$ and the measure  $d_i(s)$ for bottlenose (left) and striped dolphin (right). Panel (a) refers to the distance from the coastline $d_1(s)$, panel (b) to the vessel log-density $d_2(s)$ and panel (c) to the log-intensity from all obsevation in the SM dataset.  The lines indicate the posterior median of the estimated detection function using model \ref{eq:detdistcoast} (a), model \ref{eq:detvessel} (b) and model \ref{eq:detanimals} (c).}%
    \label{supp_fig:detection}%
\end{figure}

In Fig. \ref{supp_fig:detection}(a) (dots) we observe a considerable decrease of $q(s_{d_1})$ from 0 to about 30 Km for both species. 
The decrease is steeper for the bottlenose but clearly present also for the striped dolphin. 

Figure \ref{supp_fig:detection}(b) (dots) shows how the detection probability $q(s_{d_2})$ increases when the log-density of the vessels from the EMODnet data increases. The pattern is similar for both species.

Finally, Figure \ref{supp_fig:detection}(c) (dots) shows how the detection probability $q(s_{d_3})$ also increases when the  log-intensity estimated from all observations in the SM dataset (see \ref{supp_section:logintensity} for details) increases. 

The estimated detection functions from our models are reported in Figure \ref{supp_fig:detection}(a-c) as a solid line. They seem to capture reasonably well the observed patterns for both species. The estimated parameters for the detection functions in Equations \ref{eq:detferry}-\ref{eq:detanimals} with posterior mean and 95\% credible intervals are reported in Table \ref{tab:StenellaModel} and 
\ref{tab:TursiopeModel} for bottlenose and striped dolphin respectively.

\begin{table}[ht]
\centering
\begin{tabular}{lrrr}
  \hline
  Parameter & mean & 2.5$\%$ & 97.5$\%$ \\ 
  \hline
  \multicolumn{4}{c}{(a) Model with constant detection}\\\hline
  scale parameter $\xi_2$, Eq.~\ref{eq:detferry} & 0.66 & 0.57 & 0.75 \\ 
   \hline
  \multicolumn{4}{c}{(b) Model with detection coastline}\\\hline
  scale parameter $\xi_2$ , Eq.~\ref{eq:detferry}  & 0.56 & 0.49 & 0.62 \\ 
  scale parameter $\xi_{3,1}$, Eq.~\ref{eq:detdistcoast}   & 10.94 & 9.58 & 12.24 \\ 
   \hline
  \multicolumn{4}{c}{(d) Model with detection vessels}\\\hline
  scale parameter $\xi_2$ , Eq.~\ref{eq:detferry}  & 0.94 & 0.83 & 1.04 \\ 
  scale parameter $\xi_{3,2}$, Eq.~\ref{eq:detvessel}  & 0.60 & 0.44 & 0.74 \\ 
   location parameter $\mu_{3,2}$, Eq.~\ref{eq:detvessel}  & -2.19 & -2.50 & -1.93 \\ 
   \hline
  \multicolumn{4}{c}{(c) Model with detection intensity}\\\hline
  scale parameter $\xi_2$, Eq.~\ref{eq:detferry}  & 0.93 & 0.83 & 1.03 \\ 
  scale parameter $\xi_{3,3}$ , Eq.~\ref{eq:detanimals} & 1.07 & 0.92 & 1.21 \\ 
  location parameter $\mu_{3,3}$ , Eq.~\ref{eq:detanimals}  & 2.96 & 2.41 & 3.53 \\ 
   \hline
\end{tabular}
\caption{Model estimates for the Striped dolphin with different detection functions.}\label{tab:StenellaModel}
\end{table}

\begin{table}[ht]
\centering
\begin{tabular}{lrrr}
   \hline
  Parameter & mean & 2.5$\%$ & 97.5$\%$ \\ 
  \hline
  \multicolumn{4}{c}{(a) Model constant detection}\\
  scale parameter $\xi_2$ , Eq.~\ref{eq:detferry}  & 0.91 & 0.69 & 1.10 \\ 
     \hline
  \multicolumn{4}{c}{(b) Model with detection  coastline}\\
  scale parameter $\xi_2$, Eq.~\ref{eq:detferry}    & 0.71 & 0.57 & 0.84 \\ 
   scale parameter $\xi_{3,1}$ , Eq.~\ref{eq:detdistcoast}  & 6.71 & 5.87 & 7.52 \\ 
  
     \hline
  \multicolumn{4}{c}{(d) Model with detection vessels}\\
  scale parameter $\xi_2$, Eq.~\ref{eq:detferry}    & 0.73 & 0.59 & 0.86 \\ 
  scale parameter $\xi_{3,2}$, Eq.~\ref{eq:detvessel}  & 1.49 & 1.17 & 1.81 \\ 
  location parameter $\mu_{3,2}$ , Eq.~\ref{eq:detvessel}  & -1.78 & -2.05 & -1.52 \\ 
  \hline
  \multicolumn{4}{c}{(c) Model with detection intensity}\\
  scale parameter $\xi_2$ , Eq.~\ref{eq:detferry}  & 0.65 & 0.53 & 0.76 \\ 
   scale parameter $\xi_{3,3}$, Eq.~\ref{eq:detanimals}  & 1.08 & 0.92 & 1.24\\ 
  location parameter $\mu_{3,3}$, Eq.~\ref{eq:detanimals}  & 2.01 & 1.53 & 2.53 \\ 
  \hline
\end{tabular}
\caption{Model estimates for the Bottlenose dolphin with different detection functions}\label{tab:TursiopeModel}
\end{table}

\end{document}